\documentclass{ws-p9-75x6-50-mod}

\usepackage[reqno,centertags,intlimits]{amsmath}
\usepackage{amsfonts,amsxtra}
\usepackage{graphics}
\usepackage{openbib,cite,ifthen}

\usepackage{BQM}

\begin{document}


\title{Boundary Quantum Mechanics%
\thanks{in {\em Gravitation: Following the Prague Inspiration
{\scriptsize (To celebrate the 60th birthday of Ji\v{r}\'{\i} Bi\v{c}\'{a}k)}},
O. Semer\'{a}k, J. Podolsk\'{y}, M. \v{Z}ofka (eds.),
World Scientific, Singapore, 2002, pp. 289--323}%
}

\author{Pavel Krtou\v{s}\thanks{e-mail: {\tt Pavel.Krtous@mff.cuni.cz}}}
\address{
  Institute of Theoretical Physics,\\
  Faculty of Mathematics and Physics, Charles University,\\
  V Hole\v{s}ovi\v{c}k\'{a}ch 2, 180 00 Prague 8,\\
  Czech Republic
  }

\date{October 25, 2001}

\maketitle

\abstracts{ A reformulation of a physical theory in which measurements at the
initial and final moments of time are treated independently is discussed, both
on the classical and quantum levels. Methods of the standard quantum mechanics
are used to quantize \defterm{boundary phase space} to obtain \defterm{boundary
quantum mechanics} --- a theory that does not depend on the distinction between
the initial and final moments of time, a theory that can be formulated without
reference to the causal structure. As a supplementary material, the geometrical
description of quantization of a general (e.g. curved) configuration space is
presented.}



\section*{Introduction}
\label{sc:Intro}

\subsection*{Motivation}

In this work we formulate \defterm{boundary quantum mechanics}
--- a modification of the standard quantum mechanics where states
at the initial and final moments of time are treated independently.

Our primary motivation for building the boundary quantum mechanics was an
observation that in a field theory one usually has to define similar structures
(such as e.g., Fock bases, vacuum states, coherent states, field and momentum
observables) both for the initial and final moments of time and these structures
only differ in the time at which they are defined. An effort to
simplify handling of different structures at the initial and final time led to
a unified picture in which both time boundaries play a completely equivalent role.

The main idea is to treat both the initial and final states of a physical system in
an independent way, as if they would be states of different systems.
The Hilbert space of
boundary quantum mechanics is thus given as a tensor product of the initial and
final Hilbert spaces. The dynamics of the boundary
quantum mechanics is given by specifying a special \defterm{physical
state $|phys)$} that contains all information about dynamical correlations.

Surprisingly, the resulting theory does not need to distinguish between the initial
and final states --- it can be formulated in a way in which we do not need to
split the boundary of a (space)time domain on which we study the system to the initial
and final parts. The consequence of this fact is that we can use the formalism
of boundary quantum mechanics in situations when we do not have a reasonable
causal structure which would allow us to identify the \vague{initial} and
\vague{final} moments of time, particularly, we can use the formalism in the
Euclidian form of a theory.

The material presented here is based on some parts of Ref.~\cite{Krtous:thesis}.
The work \cite{Krtous:thesis} was concerned mainly with a study of the relation between
quantizations of a scalar field theory and relativistic particle theory.
Boundary quantum mechanics was used for an alternative description of
the scalar field theory. The general framework leading to boundary quantum
mechanics was scattered in several places in
Ref.~\cite{Krtous:thesis}. Here, the material is presented in a modified and more
compact form with an emphasis on the construction of the general formulation
of the boundary mechanics.

After boundary quantum mechanics was constructed and used in
Ref.~\cite{Krtous:thesis} the author has found out that similar ideas have been pursued in the
context of quantum gravity \cite{Ishametal:1998,Savvidou:thesis}, however, in these works the idea of
treating the different moments of time independently has been extended much
farther --- here \vague{all moments of time} are treated independently, hence,
the constructed Hilbert space of the theory is given by some kind of a continuous
tensor product of Hilbert spaces for each time.

\subsection*{Plan of work}

The explanation of the formalism is divided to three chapters.
In Chapter~\ref{sc:CTGen} a classical analogue of boundary quantum mechanics is constructed,
namely the boundary phase space is defined and its connection to standard phase
spaces is studied. This chapter formulates the theory on a very general level.
It allows to define the boundary phase space without a reference to the causal
structure. However, these details are not necessary for the following chapters ---
if one is not interested in the causal (in)dependence of the boundary phase space,
it is sufficient to understand the boundary phase space on the level
explained in the overview and summary of Chapter~\ref{sc:CTGen}.
The way of treating the general dynamical theory has been mainly inspired by
Ref.~\cite{DeWitt:book1965}.

The material presented in Chapter~\ref{sc:Quant} is actually independent of
the main subject --- of boundary quantum mechanics. In this chapter we
present a geometrical formulation of quantization for a system with a phase space
with cotangent bundle structure $\tens^\dual\valspc$ over some configuration
space $\valspc$. This is a generalization of the standard quantum mechanical
methods of quantization of the position and momentum variables to the situation when
the space of \vague{positions} $\valspc$ does not have a linear structure --- when it
is a general manifold. Only a construction of \vague{position} and \vague{momentum}
observables and their position representation is presented here, dynamical questions
are not discussed.

Finally, boundary quantum mechanics is formulated in
Chapter~\ref{sc:BQMGen}. First, the theory is built on the basis of the usual
quantum mechanics. Afterwards, it is shown that boundary
quantum mechanics can be constructed immediately from the boundary phase space without
reference to quantum mechanics at a given time. At the end the issue of dynamics is
discussed.

The main text is supplemented by three appendices in which some of the geometrical
notions are reviewed.
Appendix~\ref{apx:Dens} defines the notion of densities of
a general (complex) weight on a manifold.
Appendices~\ref{apx:SySp}, \ref{apx:TSpG} contain a
geometrical formultaion of symplectic geometry.
Most of this material is well known
(see, e.g., \cite{Arnold:book,Frankel:book})
and it is included mainly to fix the notation and remind the reader of
properties of different introduced objects.
However, let us note that Appendix~\ref{apx:TSpG} also defines
\defterm{covariant partial derivatives} on tangent and cotangent bundles ---
a notion which, to the author's knowledge, is not defined elsewhere.

\subsection*{Notation}

We use \defterm{abstract indices} to denote the tensor structure of different
tensor objects (see e.g. \cite{PenroseRindler:book}). They indicate
which space the object is from and allow us to write down a contraction in the
tensor algebra by the usual repetition of the indices. We distinguish
tensors with abstract indices from their coordinate representation.
We use bold letters for the abstract indices and normal letters for
coordinate indices (but you can hardly find them here). Hence,
choosing a base $e^{\absidx{a}}_a$ and dual base $\epsilon_{\absidx{a}}^a$,
$a=1,2,\dots$, we can write
$A^{\absidx{a}\dots}_{\absidx{b}\dots} =
  A^{a\dots}_{b\dots} \;
  e^{\absidx{a}}_{a}\,\epsilon_{\absidx{b}}^{b} \dots$,
and
$A^{a\dots}_{b\dots} =
  A^{\absidx{a}\dots}_{\absidx{b}\dots} \;
  \epsilon_{\absidx{a}}^{a} e^{\absidx{b}}_{b} \dots$.
Here $A^{\absidx{a}\dots}_{\absidx{b}\dots}$ is a tensor object and
$A^{a\dots}_{b\dots}$ is a \vague{bunch} of numbers depending on the base.

Because it can be tiresome to write indices all the time (but it is sometimes
inescapable), we drop them if it is clear what structure the object has. (In
fact, we view the abstract indices as some kind of a \vague{dress} of the tensor object
which serves to specify tensor operations.) We also use an alternative
notation for contraction using an infix operator \defterm{dot} (we use
different dots for different spaces), i.e., for example,
$a\sint \omega = \omega \sint a = a^{\absidx{n}}\omega_{\absidx{n}}$
or $(a\sint g)_{\absidx{n}} =  a^{\absidx{m}} g_{\absidx{mn}}$.


\section{Boundary, canonical, and covariant phase spaces}
\label{sc:CTGen}

\subsection*{Overview}

The main goal of this chapter is to define a boundary phase space --- a
kinematical area of the classical counterpart of boundary quantum mechanics. We
will start our construction on a very general level and we will see that the
boundary phase space can be introduced for a very broad class of theories.
However, after this general introduction we turn to a more specific theory to
grasp the meaning of the boundary phase space and to understand its relation to
standard phase spaces used in physics. We represent it as a cotangent bundle
over the boundary value space and, at the end, we introduce special types of
observables that will be quantized in the next chapter.

Before turning to a discussion of a general situation let us note that the
basic idea lying behind the boundary phase space is very simple. The boundary
phase space is a space of canonical data (\vague{values} and \vague{momenta})
at both the initial and final time with a symplectic structure induced from
canonical phase spaces at the initial and final moment of time. The main
nontrivial output of the general discussion below is a construction of the
boundary phase space without reference to the causal structure, without
necessity of splitting the boundary data to the initial and final parts.

\subsection*{The space of histories and the action}

A physical theory can be specified by a \defterm{space of elementary histories
$\histset$} and a dynamical structure on it. Elementary histories represent a
wide class of potentially imaginable evolutions of the system, not necessarily
realized in the nature.

Examples of the spaces of histories are the space of all possible trajectories
in the spacetime (theory of a relativistic particle), the space of all possible
field configuration on the spacetime (field theory), the space of all possible
connections on a spacetime (gauge field theory) or the space of all maps from
one manifold to another (target) manifold (strings, membranes, ...). The space
of histories of a general nonrelativistic system is a space of trajectories in
a configuration space $\valspc_\oix$ --- in space of \vague{positions}.

We restrict ourselves to theories that are local on some \defterm{inner
manifold $\imfld$} and we pay attention to this dependence. Generally,
histories of such local theories can be represented as sections of some fibre
bundle over the inner manifold. We use ${\scriptstyle \aixhist{x}},
{\scriptstyle \aixhist{y}}, \dots$ as tensor indices for objects from tangent
tensor spaces $\tens\,\histset$ and the dot $\ictr$ for contraction in these
spaces. Let us note that \vague{vectors} from $\tens\,\histset$ are again
sections of some bundles over the inner manifold, i.e., they are essentially
functions (or distributions), and their tensor indices ${\scriptstyle
\aixhist{x}}, \dots$ denote also the inner manifold dependence. The contraction
$\ictr$ thus includes integration over the inner manifold.

Almost all known theories can be reformulated in this way. For example,
elementary histories of a general nonrelativistic system can be viewed as
mapping from a one dimensional inner manifold $\imfld$ (\vague{time-line}
manifold) to the so-called \defterm{target manifold $\valspc_\oix$}, i.e., as
sections of the trivial fibre bundle $\imfld\times\valspc_\oix$ over $\imfld$.
The realization of a field theory is even more straightforward --- the inner
manifold is spacetime and histories are sections of some bundle over it. We
will restrict ourselves mainly to these two cases. Typical examples are a
particle in a curved space and the scalar field theory (see
\cite{Krtous:thesis} for a discussion of the latter case).

We assume that we are able to restrict the theory to any domain $\Omega$ in the
inner manifold $\imfld$. It means that we are able to speak about space of
histories $\histset[\Omega]$ on the domain $\Omega$. We will see that the
domain of dependence plays an important role in the dynamics.

On the general level, we admit any sufficiently bounded domain $\Omega$ with a
smooth boundary $\bound\Omega$. We need to deal with a \emph{bounded} domain to
assure that the action functional is well defined on a sufficiently wide set of
histories. Generally, if the domain is compact, the action is defined for all
smooth histories. However, we can also allow domains that are not compact
\vague{in some insignificant directions}. A typical example is a sandwich
domain in a globally hyperbolic spacetime between two non-intersecting
non-compact Cauchy surfaces. Such a domain is unbounded in the spatial
direction and this fact has to be compensated by a restriction of the set of
histories to those that fall-off sufficiently fast at spatial infinity. We
cannot do the same thing in the temporal direction because we would exclude
physically interesting histories --- specifically, the solutions of the
classical equations of motion. In case of a nonrelativistic system the domain
$\Omega$ is simply a compact interval in the one dimensional inner manifold
$\imfld$.

The localization of histories on the domain $\Omega$ also gives us a
localization of elements of the tangent spaces $\tens\,\histset$, i.e., we can
speak about a space $\tens\,\histset[\Omega]$. We call these tangent vectors
\defterm{linearized histories}. As we said, the tangent space at
$\hist{h}$ can be represented as a vector bundle over the inner manifold (or over
the domain $\Omega$).

The dynamics of the system is given by a domain-dependent \defterm{action
$S[\Omega]$}
\begin{equation}
  S[\Omega]\;:\;\histset\rightarrow\realn\period
\end{equation}
Let us note that we cannot generalize the action to a functional $S[\imfld]$
on the whole inner manifold --- it would be infinite for most physically
interesting histories.

The action is local, i.e., for any $\Omega$
\begin{equation}
  S[\Omega](\hist{h}_1) = S[\Omega](\hist{h}_2)
  \qquad\text{if $\hist{h}_1 = \hist{h}_2$ on $\Omega$}\commae
\end{equation}
and additive under smooth joining of domains, i.e.,
\begin{equation}
  S[\Omega](\hist{h}) =
  S[(\Omega_1](\hist{h}) + S[\Omega_2](\hist{h})\commae
\end{equation}
where $\Omega = \Omega_1 \cup \Omega_2$ is a domain and $\Omega_1 \cap
\Omega_2$ is a submanifold without boundary which is a subset of both
$\bound\Omega_1$ and $\bound\Omega_2$.

\subsection*{The equation of motion}

In general, we work with smooth histories and smooth domains with a boundary,
unless stated otherwise. We assume sufficient smoothness of the action but we
skip the discussion of this issue.

However, we explicitly assume that the action is essentially of the
first-order. In short, this means that the action leads to second-order
equations of motion. On a general level, this can be formulated by a condition
that the variation of the action (keeping the domain $\Omega$ fixed) can be
written in the following way
\begin{equation}\label{PerPartesGen}
  \grad S[\Omega](\hist{h}) =
  \charfc[\Omega]\fvaract(\hist{h}) -
  \mombndt[\bound\Omega](\hist{h})\period
\end{equation}
This relation represents an \vague{integration by parts} usually employed
in the variation of the action. A description of individual terms follows.

We use the gradient operator $\grad_{\aixhist{x}}$ on the space of histories
$\histset$ to denote the variation. It is defined by the usual relation
\begin{equation}\label{vardef}
  \variat\hist{h} \ictr \grad S[\Omega](\hist{h}) =
  \frac{d}{d\varepsilon}S[\Omega](
  \hist{h}_\varepsilon)\Big|_{\varepsilon=0}\commae
\end{equation}
where $\hist{h}_\varepsilon$ is a curve in $\histset$ with a tangent vector
$\variat\hist{h}^{\aixhist{x}}$ and, as mentioned above, the contraction
$\ictr$ also includes integration over the domain $\Omega$ in the inner
manifold.

$\fvaract_{\aixhist{x}}$ represents the \defterm{variation of the action} on
the entire inner manifold.\footnote{
  The symbol \vague{$\variat$} does not represent any operation here;
  it should be understood as a part of the symbol $\fvaract$.}
It is a 1-form on the space of histories $\histset$ and we will refer to its
tensor index $\scriptstyle{\aixhist{x}}$ as to a \defterm{variational
argument}. We require that $\fvaract(\hist{h})$ contain at most second-order
inner space derivatives of the history $\hist{h}$ and it is smooth in the
variational argument. Thanks to this smoothness, multiplication by
$\charfc[\Omega]$ is well defined.

$\charfc[\Omega]$ is the characteristic function of the domain $\Omega$ (i.e.,
$\charfc[\Omega]=1$ inside $\Omega$ and zero outside). We will also use a
bi-distribution $(\charfc[\Omega]\deltadst)$ which projects (\vague{cuts-off})
smooth functions to functions with a support on $\Omega$ (cf.
Appendix~\ref{apx:Dens}).

$\mombndt_{\aixhist{x}}[\bound\Omega]$ is the \defterm{generalized momentum on
the boundary $\bound\Omega$}. It is localized on the boundary in both its
history dependence and in the variational argument (the momentum is a 1-form on
$\histset$ and by variational argument we again mean the
${\scriptstyle{\aixhist{x}}}$-dependence). We require that it can contain at
most the first derivative in the direction normal to the boundary and cannot
contain any normal derivatives in the variational argument. Hence, it can be
represented as a distribution on boundary values of the linearized histories.
Later (cf. Eqs.~\eqref{genmomnrel} and \eqref{genmomDef}) we clarify a relation
of the generalized momentum to the usual definition of the momentum in the
Lagrangian formalism.

The \defterm{classical equations of motion} are given by the condition
\begin{equation}\label{EqOfMotGen}
  \fvaract(\hist{h}) = 0 \commae
\end{equation}
and we denote the space of their solutions by $\sphspc$ --- it is the \defterm{space
of classical solutions}.

The \defterm{linearized equation of motion} selects the linearized histories
tangent to $\sphspc$. It has the form
\begin{equation}\label{lineqmot}
  \svaractr(\hist{h}) \ictr \variat \hist{h} = 0 \commae
\end{equation}
where $\variat \hist{h}$ is a linearized history (tangent vector to $\histset$)
at a classical history $\hist{h}$ and the \defterm{second variation of the
action $\svaractr_{\aixhist{xy}}$}, given by
\begin{equation}\label{svaractDef}
  \svaractr_{\aixhist{xy}}(\hist{h}) =
  \histcnx_{\aixhist{y}} \fvaract_{\aixhist{x}}(\hist{h})\commae
\end{equation}
is a derivative of the form $\fvaract$ using an ultralocal connection
$\histcnx$. It is easy to check that for a classical history $\hist{h}$, thanks
to \eqref{EqOfMotGen}, the second variation of the action $\svaractr(\hist{h})$
does not depend on the choice of the connection $\histcnx$. We define
$\svaractl_{\aixhist{xy}}=\svaractr_{\aixhist{yx}}$ and
\begin{equation}
  \svaractr[\Omega] = (\charfc[\Omega]\deltadst)\ictr\svaractr \comma
  \svaractl[\Omega] = \svaractl\ictr(\charfc[\Omega]\deltadst) \commae
\end{equation}
satisfying again $\svaractl_{\aixhist{xy}}[\Omega]
= \svaractr_{\aixhist{yx}}[\Omega]$.

We assume that the equation of motion has a well-defined boundary problem on
the domain $\Omega$, i.e., we assume the existence of a unique solution from
$\sphspc$ for a given restriction of a history to the boundary. Moreover, we
assume that the linearized equation of motion has a well-formulated Dirichlet
and Neumann boundary problem, i.e., there is a unique solution to the
linearized equation of motion for a given linearized value on the boundary or
linearized momentum on the boundary. This requires some generality of the
action --- for example we exclude a massless scalar field.  More serious is the
restriction that we must also exclude theories with local symmetries --- see
\cite{DeWitt:book1965} for some details on this case.

When working with the manifold $\sphspc$, we use $\aixsph{A}, \aixsph{B},
\dots$ for tangent tensor indices.

\subsection*{Boundary phase space}

Next we define the \defterm{boundary symplectic structure $\grad
\mombndt[\bound\Omega]$} to be the external derivative of the generalized
momentum 1-form $\mombndt[\bound\Omega]$, which turns out to be the Wronskian
of the second variation of the action (see~\eqref{svaractDef}),
\begin{equation}\label{biFcDefGen}
  \grad\mombndt[\bound\Omega] =
  \svaractl[\Omega] - \svaractr[\Omega]\period
\end{equation}

We say that two histories are canonically equivalent on the boundary if they
have the same restriction on the boundary and the same momentum $\mombndt$. We
call the quotient of the space $\histset$ with respect to this equivalence the
\defterm{boundary phase space $\bphspc[\bound\Omega]$}. A point from the
boundary phase space thus represents values and momenta on the entire boundary
$\bound\Omega$, i.e., at \emph{both} the initial and final time.

We use $\aixbph{A},\aixbph{B},\dots$ as tensor indices for tensors from the tangent
spaces $\tens\,\bphspc[\bound\Omega]$.

It is straightforward to check that vectors tangent to the orbits of
equivalence are degenerate directions of the boundary symplectic form
$\grad\mombndt[\bound\Omega]$ and therefore we can define its action on the
space $\bphspc[\bound\Omega]$. We will require that the form
$\grad\mombndt[\bound\Omega]$ is non-degenerate on the boundary phase space.
Because the external derivative of this form is zero, it is indeed a symplectic
form in the sense of Appendix \ref{apx:SySp}, and it gives a symplectic space
structure to the space $\bphspc[\bound\Omega]$, thus justifying the name
boundary \emph{phase} space.

The space $\sphspc$ is a submanifold of $\histset$ and, therefore, it defines a
submanifold of the space $\bphspc[\bound\Omega]$, which we denote by the same
letter $\sphspc$.

\subsection*{Lagrangian density}
\label{ssc:LagrDens}

Until now, we have been developing the formalism on a very general level. In the
following we restrict to theories with the action given by
\begin{equation}
  S[\Omega](\hist{h}) = \int_\Omega \lagrdst(\hist{h},\Dmap \hist{h}) \commae
\end{equation}
where $\lagrdst$ is the \defterm{Lagrangian density} --- density on the inner
manifold $\imfld$ --- ultralocally dependent on the value of the history and
\vague{velocities}, i.e., inner space derivatives of the history.\footnote{
  The formalism developed until now is more general --- it covers for example the
  case of the Einstein-Hilbert action for gravity for which the Lagrangian
  density contains second spacetime derivatives of the metric. But this
  dependence is degenerate and it is possible to satisfy the above conditions
  if a proper boundary term is chosen.}

In general, we need some additional structure on the fibre bundle $\histset$ to
define the \vague{velocity} (an inner space derivative of the history) --- we
need, for example, a connection on the bundle. There can exist a natural
connection (e.g. if we can identify fibres of the bundle) or a choice of the
connection can be equivalent to a specification of an external field (a gauge
field of Yang-Mills theories). In the following we will have in mind mainly a
general nonrelativistic system, where we do not need any additional structure
--- the velocity can be simply understood as a derivative of the trajectory
$\hist{h} : \imfld\rightarrow\valspc_\oix$ with respect to the \vague{proper
time} (a preferred coordinate on $\imfld$).

In this simple case the variation of the action
and an integration by parts gives us the decomposition \eqref{PerPartesGen} with
$\fvaract$ and $\mombndt[\bound\Omega]$ as follows:
\begin{gather}
  \fvaract (\hist{h})=
  \frac{\valcnx\lagrdst}{\partial \hist{h}}
  (\hist{h},\dot{\hist{h}})
  - {\Bigl(\frac{\partial\lagrdst}
  {\partial\dot{\hist{h}}}
  (\hist{h},\dot{\hist{h}})\Bigr)}\spdot \commae\\
  \mombndt[\bound\Omega](\hist{h}) =
  \frac{\partial\lagrdst}{\partial\dot{\hist{h}}}
  (\hist{h},\dot{\hist{h}})\,
  \deltafc[\bound\Omega] \period\label{genmomnrel}
\end{gather}
Here $\deltafc[\bound\Omega]$ is a delta function localized on the boundary
$\bound\Omega$, i.e., with a support at the end points of the time interval
$\Omega$. A similar expression for the general case when the inner manifold is
not one dimensional and a thorough discussion of the scalar field case can be
found in \cite{Krtous:thesis}.

\subsection*{Space of boundary values}

We call the restriction of the history to the boundary $\bound\Omega$ the
\defterm{boundary value} of the history. We emphasize that the
boundary values do not include any inner space derivatives of the histories in
the direction normal to the boundary. The \defterm{space of all boundary
values} $\valspc[\bound\Omega]$ can be represented as a fibre bundle with the
boundary $\bound\Omega$ as the base manifold. We use ${\scriptstyle\aixval{x}},
{\scriptstyle\aixval{y}},\dots$ as tensor indices for tensors from the tangent
spaces $\tens\,\valspc[\bound\Omega]$ and the dot $\sint$ for contraction in
these spaces. In general, the boundary $\bound\Omega$ is a manifold and the
contraction $\sint$ includes integration over the boundary.

Let us note that the choice of notation is adjusted to the scalar field theory.
In this case the inner manifold is spacetime, the target manifold are real
numbers~$\realn$, the space of histories $\histset$ is the space of functions
on spacetime, and the indices ${\scriptstyle \aixhist{x}}, \dots$ used for
vectors from $\tens\,\histset$ actually represent points in spacetime.
Similarly, the space $\valspc[\bound\Omega]$ is the space of functions on the
boundary $\bound\Omega$ of a spacetime domain $\Omega$ and indices
${\scriptstyle\aixval{x}}, \dots$ represent points on the boundary. In case of
the target manifold being not so simple, all these indices also carry
information about a direction in the target manifold.

Another simple case is the nonrelativistic system where the inner manifold is
one dimensional. In this case the linearized history
$\variat\hist{h}^{\aixhist{x}}$ represents a time dependent
target-manifold-vector-valued function, i.e., the index ${\scriptstyle
\aixhist{x}}$ represents a time variable and a direction in the target
manifold. The boundary $\bound\Omega$ consists only of two points from $\imfld$
and the boundary value $\bhist{x}$ of a trajectory $\hist{h}$ is a pair of end
points $\bhist{x} =[\bhist{x}_\fix,\bhist{x}_\iix]$. The space of boundary
values $\valspc[\bound\Omega]$ is thus isomorfic to
$\valspc_\oix\times\valspc_\oix$ and the tangent space
$\tens\,\valspc[\bound\Omega]$ to a direct sum
$\tens\,\valspc_\oix\oplus\tens\,\valspc_\oix$. Therefore, the tensor indices
${\scriptstyle\aixval{x}}, \dots$ correspond to pairs of directions in the
target manifold and the contraction $\sint$ reduces to a contraction over
finite dimensional vector spaces.

We denote the projection from $\histset$ to $\valspc[\bound\Omega]$ by
$\genval[\bound\Omega]$. We already said that the condition that the
generalized momentum does not contain inner space derivatives normal to the
boundary in its variational argument ensures that
$\mombndt_{\aixhist{x}}[\bound\Omega]$ can be realized as a distribution (in
the ${\scriptstyle \aixhist{x}}$ argument) on the boundary values od the
linearized histories --- we do not need any other information about a
linearized history $\variat\hist{h}^{\aixhist{x}}$ except its value on the
boundary to compute
$\variat\hist{h}^{\aixhist{x}}\mombndt_{\aixhist{x}}[\bound\Omega]$.
Translation between linearized histories and its boundary values is, of course,
the differential $\Dmap\genval[\bound\Omega]$ of the projection
$\genval[\bound\Omega]$. Hence, we can represent the generalized momentum in
the following form:
\begin{equation}\label{genmomDef}
  \mombndt_{\aixhist{x}}[\bound\Omega] =
  \genmom_{\aixval{x}}[\bound\Omega]\:
  \Dmap_{\aixhist{x}}^{\aixval{x}}\genval[\bound\Omega]\period
\end{equation}
Here $\genmom[\bound\Omega](\hist{h})$ is from the cotangent bundle
$\tens_{\genval(\hist{h})}^\dual\valspc[\bound\Omega]$. The differential
$\Dmap_{\aixhist{x}}^{\aixval{x}}\genval[\bound\Omega]$ understood as a
distribution (in the ${\scriptstyle \aixhist{x}}$ argument) is actually a delta
function with a support on the boundary (multiplied by a (finite dimensional)
unit tensor on the target manifold).

$\genval[\bound\Omega](\hist{h})$ and $\genmom[\bound\Omega](\hist{h})$
represent the values and the momenta of the history $\hist{h}$ at \emph{both}
the initial and final time.  The meaning of $\genval[\bound\Omega](\hist{h})$
is clear from its definition; the meaning of $\genmom[\bound\Omega](\hist{h})$
can be seen --- at least in case of a one dimensional inner manifold --- by
comparing Eq.~\eqref{genmomDef} and Eq.~\eqref{genmomnrel}. The same
interpretation holds in the general case, too (see \cite{Krtous:thesis} for the
details of the scalar field case). In the following, we drop the boundary
dependence of $\genval$ and $\genmom$.

Next we define the \defterm{classical history $\histsol(\bhist{x})$} with a
given boundary value $\bhist{x}$
\begin{equation}\label{histsolDef}
  \fvaract(\histsol(\bhist{x})) = 0  \comma
  \genval(\histsol(\bhist{x})) = \bhist{x}
\end{equation}
and the classical action
\begin{equation}\label{classactionDef}
  \extr{S}[\Omega](\bhist{x}) =
  S[\Omega](\histsol(\bhist{x})) \period
\end{equation}

We use $\histsol$ also for the induced map from the space
$\valspc[\bound\Omega]$ to the boundary phase space $\bphspc[\bound\Omega]$,
and $\genval$ and $\genmom$ for the induced maps from the boundary phase space
$\bphspc[\bound\Omega]$ to the spaces $\valspc[\bound\Omega]$ and
$\tens^\dual\,\valspc[\bound\Omega]$. This suggests that we can represent the
boundary phase space $\bphspc[\bound\Omega]$ as a cotangent bundle
$\tens^\dual\,\valspc[\bound\Omega]$. Indeed, the canonical symplectic
structure of the cotangent bundle \eqref{CanCTBSymplstr} does coincide with
$\grad\mombndt[\bound\Omega]$:
\begin{equation}\label{dmombndtsymplstr}
  \valcnx_{\aixbph{A}}\genmom_{\aixval{x}} \wedge
  \Dmap_{\aixbph{B}}^{\aixval{x}}\genval =
  \grad_{\aixbph{A}} ( \genmom_{\aixval{x}}
  \Dmap_{\aixbph{B}}^{\aixval{x}}\genval) =
  \grad_{\aixbph{A}} \mombndt_{\aixbph{B}} \period
\end{equation}

The space $\sphspc$ as a submanifold of $\bphspc[\bound\Omega]$ can be
characterized using the condition
\begin{equation}\label{SphspcInBphspc}
  \genmom = - \grad \extr{S} (\genval) \commae
\end{equation}
which follows from
\begin{equation}
\begin{split}
  \grad_{\aixval{x}}\extr{S}
  &=\Dmap_{\aixval{x}}^{\aixhist{x}}\histsol\;
    \grad_{\aixhist{x}}S(\histsol) =
    \Dmap_{\aixval{x}}^{\aixhist{x}}\histsol\;
    \bigl(\fvaract_{\aixhist{x}}[\Omega](\histsol) -
    \mombndt_{\aixhist{x}}[\bound\Omega](\histsol)\bigr) = \\
  &=-\Dmap_{\aixval{x}}^{\aixhist{x}}\histsol\;
    \Dmap_{\aixhist{x}}^{\aixval{y}}\genval(\histsol)\;
    \genmom_{\aixval{y}}(\histsol) =
    -\genmom_{\aixval{x}}(\histsol) \period
\end{split}
\end{equation}
Linearization of Eq. \eqref{SphspcInBphspc} gives
\begin{equation} \label{DHistsol}
  \Dmap_{\aixval{x}}^{\aixbph{A}} \histsol =
  \frac{\valcnx_{\aixval{x}}^{\aixbph{A}}}{\partial \genval} -
  \bigl(\valcnx_{\aixval{x}}\grad_{\aixval{y}}\extr{S}\bigr)
  \frac{\partial^{\aixbph{A}}}{\partial \genmom_{\aixval{y}}}
\end{equation}
(see Appendix~\ref{apx:TSpG} for definitions of objects used here, especially
Eq.~\eqref{ctgdual}).

\subsection*{Causal structure}

Until now we have not needed any time flow in the underlying inner manifold
$\imfld$. It could be spacetime, an inner sheet of a string, or a one
dimensional time-line --- in all these cases we do have some kind of time flow.
However, the formalism also works in a more general situation.  We can use it,
for example, for the Euclidian form of the theory, where we do not have any
time direction. Now, we will use the time flow for the first time and we will
add an additional causal structure that will allow us to define concepts such
as canonical and covariant phase spaces.

All what we will use is an assumtion that the boundary of the domain can be
split into two disjoint parts without a boundary
\begin{equation}\label{CausDecomp}\begin{gathered}
  \bound\Omega = \bound\Omega_\fix \cup \bound\Omega_\iix =
  - \Sigma_\fix \cup \Sigma_\iix \commae\\
  \bound\Omega_\fix = - \Sigma_\fix \comma
  \bound\Omega_\iix = \Sigma_\iix\commae
\end{gathered}\end{equation}
each of them carrying a full set of data (see the condition below). Here the
minus sign suggests an opposite choice of the normal direction orientation for
one part of the boundary. Clearly, we have in mind two Cauchy hypersurfaces
that define a sandwich domain in a globally hyperbolic spacetime, or two end
points of the interval in the one dimensional inner manifold $\imfld$ in the
case of a non-relativistic system. The decomposition in \eqref{CausDecomp}
allows us to write
\begin{equation}\begin{gathered}\label{ifSpcDecomp}
  \valspc[\bound\Omega] =
  \valspc[\Sigma_\fix] \times \valspc[\Sigma_\iix] \commae\\
  \bphspc[\bound\Omega] =
  - \bphspc[\Sigma_\fix] \oplus \bphspc[\Sigma_\iix] \commae\\
  \tens\,\valspc[\bound\Omega] =
  \tens\,\valspc[\Sigma_\fix] \oplus
  \tens\,\valspc[\Sigma_\iix] \commae
\end{gathered}\end{equation}
and we will use shorthands $\valspc$, $\valspc_\fix$, $\valspc_\iix$ and
$\bphspc$, $\bphspc_\fix$, $\bphspc_\iix$.

In case of field theory, the space $\valspc_\fix$ (or $\valspc_\iix$,
respectively) represents the space of field configurations on the final (or the
initial) Cauchy hypersurface. In case of particle theory, spaces
$\valspc_\fix$, $\valspc_\iix$ represent positions of the particle at the final
or initial time, i.e., $\valspc_\fix=\valspc_\iix=\valspc_\oix$. Similarly
$\bphspc_\fix$ (or $\bphspc_\iix$) represent canonical data (values and
momenta) at the final (or initial) time.

We require that both parts contain a full set of boundary data --- there should
exist a unique classical history for a given element from $\bphspc_\fix$ or
$\bphspc_\iix$.

Thanks to the locality, we can decompose the symplectic structure
$\grad\mombndt[\bound\Omega]$ as
\begin{equation}\label{fisymplstrdecomp}
  \grad\mombndt[\bound\Omega] =
  - \grad\mombndt[\Sigma_\fix] + \grad\mombndt[\Sigma_\iix] \period
\end{equation}

$\grad\mombndt[\Sigma_\fix]$ and $\grad\mombndt[\Sigma_\iix]$ play the role of
the symplectic structure on $\bphspc_\fix$ and $\bphspc_\iix$. We will call
these spaces \defterm{canonical phase spaces}. The minus sign in the relations
\eqref{ifSpcDecomp} reflects the relation of these spaces as symplectic spaces.

The canonical phase spaces can be again represented as cotangent bundles
$\tens\,\valspc_\fix$ and $\tens\,\valspc_\iix$ through the maps
$\genval_\fix$, $\genmom_\fix$ and $\genval_\iix$, $\genmom_\iix$. Let us note
that $\genmom_\fix$ takes into account the opposite orientation of the normal
direction to $\Sigma_\fix$ and $\bound\Omega_\fix$, so that
\begin{equation}\label{fidecompofp}
  \genmom = - \genmom_\fix \oplus \genmom_\iix\period
\end{equation}

\subsection*{Covariant phase space}

Finally, we can also give a phase space structure  to the space of classical
histories $\sphspc$. First we note that, thanks to \eqref{biFcDefGen},
solutions $\xi_1,\xi_2\in \tens\,\sphspc$ of linearized equations of motion
\eqref{lineqmot} satisfy
\begin{equation}
  \xi_1\ictr\grad\mombndt[\bound\Omega]\ictr\xi_2 = 0\period
\end{equation}
Therefore, it follows from \eqref{fisymplstrdecomp} that
$\grad\mombndt[\Sigma_\fix]$ and $\grad\mombndt[\Sigma_\iix]$ have
the same restriction $\symplstr$ on the space $\sphspc$.
\begin{equation}
  \xi_1^{\aixsph{A}}\symplstr_{\aixsph{AB}}\xi_2^{\aixsph{B}} =
  \xi_1\ictr\grad\mombndt[\Sigma_\fix]\ictr\xi_2 =
  \xi_1\ictr\grad\mombndt[\Sigma_\iix]\ictr\xi_2 \period
\end{equation}
In the same way, we check that the same expression for $\symplstr$ holds for
any future-oriented Cauchy hypersurface $\Sigma$. It means that we have
equipped the space of classical histories $\sphspc$ with the symplectic
structure $\symplstr$. We will call this space the \defterm{covariant phase
space}. From Eq.~\eqref{dmombndtsymplstr} follows that the $\valspc_{\fix}$,
$\valspc_{\iix}$ and $\tens^\dual\,\valspc_{\fix}$,
$\tens^\dual\,\valspc_{\iix}$-valued observables $\genval_\fix$, $\genval_\iix$
and $\genmom_\fix$, $\genmom_\iix$ are canonically conjugate on this space (cf.
Appendix~\ref{apx:SySp}):
\begin{equation}\label{ifCanVar}
  \symplstr_{\aixsph{AB}} =
  \valcnx_{\aixsph{A}}\genmom_{\fix\aixval{x}} \wedge
  \Dmap_{\aixsph{B}}^{\aixval{x}}\genval_\fix =
  \valcnx_{\aixsph{A}}\genmom_{\iix\aixval{x}} \wedge
  \Dmap_{\aixsph{B}}^{\aixval{x}}\genval_\iix \period
\end{equation}
We can invert the symplectic form $\symplstr$ to get $\symplstr^{\dash 1}$:
\begin{equation}\label{InvSymstr}
  \symplstr^{\dash 1\,\aixsph{AM}} \symplstr_{\aixsph{BM}} =
  \symplstr^{\dash 1\,\aixsph{MA}} \symplstr_{\aixsph{MB}} =
  \deltadst_{\aixsph{B}}^{\aixsph{A}} \period
\end{equation}

If we view $\sphspc$ as a subspace of the boundary phase space $\bphspc$ we can
understand $\symplstr^{\dash 1}$ as a tensor from $\tens^2\,\bphspc$ tangent to
$\sphspc$ in both indices. Because the differential $\Dmap\histsol$ of the map
$\histsol : \valspc\rightarrow\bphspc$ (it is a restriction of the
map~\eqref{histsolDef} to $\bphspc$) plays the role of a projector of vectors
from $\tens\valspc$ to vectors from $\tens\bphspc$ tangent to $\sphspc$, we can
write
\begin{equation}
  \symplstr^{\dash 1} =
  \Dmap\histsol \sint \gFC \sint \Dmap\histsol
\end{equation}
with an antisymmetric tensor $\gFC\in\tens^2\,\valspc$. Using \eqref{DHistsol},
\eqref{InvSymstr}, and \eqref{ifCanVar} we get
\begin{equation}
  \gFC\sint (\grad_\fix\grad_\iix\extr{S} -
  \grad_\iix\grad_\fix\extr{S}) = \deltadst_\valspc \period
\end{equation}
This means that
\begin{equation}\begin{gathered}
  \gFC = \gFif - \gFfi\comma
  \gFif^{\,\aixval{yx}} = \gFfi^{\,\aixval{xy}}\commae\\
  \gFif\sint\grad_\fix\grad_\iix\extr{S} =
  \deltadst_{\valspc_\iix} \comma
  \gFfi\sint\grad_\iix\grad_\fix\extr{S} =
  \deltadst_{\valspc_\fix}  \commae
\end{gathered}\end{equation}
and, using Eq.~\eqref{DHistsol} again, we get
\begin{equation}\label{symplstrinvfi}
\begin{split}
  \symplstr^{\dash 1}
  &= \Bigl(
    \frac{\valcnx_{\aixval{x}}}{\partial\genval} -
    \frac{\partial}{\partial\genmom_{\aixval{u}}}
    (\valcnx_{\aixval{u}}\grad_{\aixval{x}}\extr{S})
    \Bigr)\:\gFC^{\aixval{xy}}\:\Bigl
    (\frac{\valcnx_{\aixval{y}}}{\partial\genval} -
    (\valcnx_{\aixval{y}}\grad_{\aixval{v}}\extr{S})
    \frac{\partial}{\partial\genmom_{\aixval{v}}}
    \Bigr) =\\
  &= \Bigl(
    \frac{\partial}{\partial\genmom_{\fix\aixval{u}}}
    \frac{\valcnx_{\aixval{u}}}{\partial\genval_\fix} -
    \frac{\valcnx_{\aixval{u}}}{\partial\genval_\fix}
    \frac{\partial}{\partial\genmom_{\fix\aixval{u}}}
    \Bigr) + \Bigl(
    \frac{\partial}{\partial\genmom_{\iix\aixval{u}}}
    \frac{\valcnx_{\aixval{u}}}{\partial\genval_\iix} -
    \frac{\valcnx_{\aixval{u}}}{\partial\genval_\iix}
    \frac{\partial}{\partial\genmom_{\iix\aixval{u}}}
    \Bigr) +\\
  &\qquad+ \Bigl(
    \frac{\valcnx_{\aixval{x}}}{\partial\genval}
    \:\gFC^{\aixval{xy}}\:
    \frac{\valcnx_{\aixval{y}}}{\partial\genval}
    \Bigr) +\\
  &\qquad+\biggl(
    \frac{\partial}{\partial\genmom_{\fix\aixval{x}}}
    \Bigl(
    (\valcnx_{\fix{\aixval{x}}}\grad_{\fix{\aixval{u}}}\extr{S})
    \gFfi^{\aixval{uv}}
    (\valcnx_{\iix{\aixval{v}}}\grad_{\iix{\aixval{y}}}\extr{S})
    - \grad_{\fix{\aixval{x}}}\grad_{\iix{\aixval{y}}}\extr{S}
    \Bigr)
    \frac{\partial}{\partial\genmom_{\iix\aixval{y}}}\\
  &\qquad\qquad\qquad -
    \frac{\partial}{\partial\genmom_{\iix\aixval{x}}}
    \Bigl(
    (\valcnx_{\iix{\aixval{x}}}\grad_{\iix{\aixval{u}}}\extr{S})
    \gFif^{\aixval{uv}}
    (\valcnx_{\fix{\aixval{v}}}\grad_{\fix{\aixval{y}}}\extr{S})
    - \grad_{\iix{\aixval{x}}}\grad_{\fix{\aixval{y}}}\extr{S}
    \Bigr)
    \frac{\partial}{\partial\genmom_{\fix\aixval{y}}}
    \biggr) + \\
  &\qquad+\Bigl(
    \frac{\valcnx_{\aixval{x}}}{\partial\genval_\iix}
    \gFif^{\aixval{xy}}
    (\valcnx_{\fix\aixval{y}}\grad_{\fix\aixval{u}}\extr{S})
    \frac{\partial}{\partial\genmom_{\fix\aixval{u}}}
    -
    \frac{\partial}{\partial\genmom_{\fix\aixval{u}}}
    (\valcnx_{\fix\aixval{u}}\grad_{\fix\aixval{y}}\extr{S})
    \gFfi^{\aixval{yx}}
    \frac{\valcnx_{\aixval{x}}}{\partial\genval_\iix}
    \Bigr) +\\
  &\qquad+\Bigl(
    \frac{\valcnx_{\aixval{x}}}{\partial\genval_\fix}
    \gFfi^{\aixval{xy}}
    (\valcnx_{\iix\aixval{y}}\grad_{\iix\aixval{u}}\extr{S})
    \frac{\partial}{\partial\genmom_{\iix\aixval{u}}}
    -
    \frac{\partial}{\partial\genmom_{\iix\aixval{u}}}
    (\valcnx_{\iix\aixval{u}}\grad_{\iix\aixval{y}}\extr{S})
    \gFif^{\aixval{yx}}
    \frac{\valcnx_{\aixval{x}}}{\partial\genval_\fix}
    \Bigr) \period
\end{split}
\end{equation}
Here, $\valcnx$ is any covariant derivative on the value space $\valspc$,
$\valcnx_{\fix}$ and $\valcnx_{\iix}$ are its restrictions to $\valspc_\fix$
and $\valspc_\iix$, and $\grad_\fix$, $\grad_\iix$ are gradients on
$\valspc_\fix$ and $\valspc_\iix$.

Or, if we view $\sphspc$ as a subspace of the space of histories $\histset$ we
can understand $\symplstr^{\dash 1}$ as a tangent tensor from
$\tens^2\,\histset$ that satisfies the linear equation of motion in both
indices. We call this representation the \defterm{causal Green function $\GFC$}
\begin{equation}\label{GFCdef}
  \GFC^{\aixhist{xy}} =
  \Dmap_{\aixval{x}}^{\aixhist{x}}\histsol\:\gFC^{\aixval{xy}}\:
  \Dmap_{\aixval{y}}^{\aixhist{y}}\histsol \commae
\end{equation}
where we now understand $\histsol$ as a map from $\valspc$ to $\histset$. In
the space $\tens\,\histset$, Eq.~\eqref{InvSymstr} takes the form
\begin{equation}
  \GFC \ictr \grad\mombndt[\Sigma] = - \Dcausal[\Sigma] \commae
\end{equation}
where $\Dcausal[\Sigma]$ is a \defterm{Cauchy projector} of a history on the
linearized classical history with the same value and momentum on the surface
$\Sigma$. It is, of course, an identity on $\tens\,\sphspc$.

Similarly to Eq.~\eqref{GFCdef} we can introduce $\GFif$ and $\GFfi$, which
turn out to be the advanced and retarded Green functions. $\gFC$, $\gFif$, and
$\gFfi$ are thus the corresponding Green functions evaluated on the boundary
$\bound\Omega$.

\subsection*{Poisson brackets}

The \defterm{Poisson brackets} of two observables on a phase space are defined
by \eqref{PoisBrDef}. We can compare Poisson brackets in the sense of different
phase spaces. Clearly, any observable on $\histset$ generates an observable on
$\sphspc$ and we can define
\begin{equation}
  \{A,B\}_\sphspc =
  \grad_{\aixhist{x}}A\: \GFC^{\aixhist{xy}}\:
  \grad_{\aixhist{y}} B
  \quad\text{on}\quad \sphspc\period
\end{equation}
For observables depending only on the boundary values and momenta --- i.e., for
observables on $\bphspc$ --- we can define the Poisson brackets in the sense of
the boundary phase space
\begin{equation}
  \{A,B\}_\bphspc =
  \grad_{\aixbph{A}}A\; \grad\mombndt^{\dash 1\,\aixbph{AB}}\;
  \grad_{\aixbph{B}} B =
  \frac{\partial A}{\partial\genmom_{\aixval{x}}}
  \frac{\valcnx_{\aixval{x}} B}{\partial\genval} -
  \frac{\valcnx_{\aixval{x}} A}{\partial\genval}
  \frac{\partial B}{\partial\genmom_{\aixval{x}}}\period
\end{equation}
With the help of \eqref{symplstrinvfi} we find that the covariant Poisson
brackets for such observables are given by
\begin{equation}
  \begin{split}\raisetag{100pt}
  \{A,B\}_\sphspc
  &=\grad_{\aixbph{A}}A\,
    \symplstr^{\dash 1\,\aixbph{AB}}\,
    \grad_{\aixbph{B}} B =\\
  &= \Bigl(
    \frac{\partial A}{\partial\genmom_{\fix\aixval{u}}}
    \frac{\valcnx_{\aixval{u}} B}{\partial\genval_\fix} -
    \frac{\valcnx_{\aixval{u}} A}{\partial\genval_\fix}
    \frac{\partial A}{\partial\genmom_{\fix\aixval{u}}}
    \Bigr) + \Bigl(
    \frac{\partial A}{\partial\genmom_{\iix\aixval{u}}}
    \frac{\valcnx_{\aixval{u}} B}{\partial\genval_\iix} -
    \frac{\valcnx_{\aixval{u}} A}{\partial\genval_\iix}
    \frac{\partial B}{\partial\genmom_{\iix\aixval{u}}}
    \Bigr) +\\
  &\qquad+ \Bigl(
    \frac{\valcnx_{\aixval{x}} A}{\partial\genval}
    \:\gFC^{\aixval{xy}}\:
    \frac{\valcnx_{\aixval{y}} B}{\partial\genval}
    \Bigr) +\\
  &\qquad+\biggl(
    \frac{\partial A}{\partial\genmom_{\fix\aixval{x}}}
    \Bigl(
    (\valcnx_{\fix{\aixval{x}}}\grad_{\fix{\aixval{u}}}\extr{S})
    \gFfi^{\aixval{uv}}
    (\valcnx_{\iix{\aixval{v}}}\grad_{\iix{\aixval{y}}}\extr{S})
    - \grad_{\fix{\aixval{x}}}\grad_{\iix{\aixval{y}}}\extr{S}
    \Bigr)
    \frac{\partial B}{\partial\genmom_{\iix\aixval{y}}}\\
  &\qquad\qquad\qquad-
    \frac{\partial A}{\partial\genmom_{\iix\aixval{x}}}
    \Bigr(
    (\valcnx_{\iix{\aixval{x}}}\grad_{\iix{\aixval{u}}}\extr{S})
    \gFif^{\aixval{uv}}
    (\valcnx_{\fix{\aixval{v}}}\grad_{\fix{\aixval{y}}}\extr{S})
    - \grad_{\iix{\aixval{x}}}\grad_{\fix{\aixval{y}}}\extr{S}
    \Bigr)
    \frac{\partial B}{\partial\genmom_{\fix\aixval{y}}}
    \biggr) +\\
  &\qquad+\Bigl(
    \frac{\valcnx_{\aixval{x}}A}{\partial\genval_\iix}
    \gFif^{\aixval{xy}}
    (\valcnx_{\fix\aixval{y}}\grad_{\fix\aixval{u}}\extr{S})
    \frac{\partial B}{\partial\genmom_{\fix\aixval{u}}}
    -
    \frac{\partial A}{\partial\genmom_{\fix\aixval{u}}}
    (\valcnx_{\fix\aixval{u}}\grad_{\fix\aixval{y}}\extr{S})
    \gFfi^{\aixval{yx}}
    \frac{\valcnx_{\aixval{x}}B}{\partial\genval_\iix}
    \Bigr) +\\
  &\qquad+\Bigl(
    \frac{\valcnx_{\aixval{x}}A}{\partial\genval_\fix}
    \gFfi^{\aixval{xy}}
    (\valcnx_{\iix\aixval{y}}\grad_{\iix\aixval{u}}\extr{S})
    \frac{\partial B}{\partial\genmom_{\iix\aixval{u}}}
    -
    \frac{\partial A}{\partial\genmom_{\iix\aixval{u}}}
    (\valcnx_{\iix\aixval{u}}\grad_{\iix\aixval{y}}\extr{S})
    \gFif^{\aixval{yx}}
    \frac{\valcnx_{\aixval{x}} B}{\partial\genval_\fix}
    \Bigr) \period
\end{split}
\end{equation}
Moreover, for observables localized only on $\Sigma_\fix$ or $\Sigma_\iix$ we
have
\begin{equation}\begin{gathered}
  \{A_\fix,B_\fix\}_{\bphspc_\fix} =
  -\{A_\fix,B_\fix\}_{\bphspc} =
  \{A_\fix,B_\fix\}_{\sphspc} \commae\\
  \{A_\iix,B_\iix\}_{\bphspc_\iix} =
  \{A_\iix,B_\iix\}_{\bphspc} =
  \{A_\iix,B_\iix\}_{\sphspc} \period
\end{gathered}\end{equation}

\subsection*{Observables at most linear in momenta}

During the quantization we will be interested in a special kind
of observables on the boundary phase space $\bphspc$ (or, in general, on any phase space
with the cotangent bundle structure). We will investigate observables
dependent only on the value and observables linear in the momentum. We define
the following observables for a function $f$ and a vector field $a$ on
$\valspc$
\begin{gather}
  \Fobs_f = f(\genval) \commae \label{FobsDef}\\
  \Gobs_a = a^{\aixval{x}}(\genval)\,\genmom_{\aixval{x}}
  \period \label{GobsDef}
\end{gather}
The Poisson brackets of these observables are
\begin{equation}\begin{aligned}\label{FGPoissBr}
  \{\Fobs_{f_1},\Fobs_{f_2}\}_\bphspc &= 0\commae\\
  \{\Fobs_{f},\Gobs_{a}\}_\bphspc &= - \Fobs_{a\sint\grad f}\commae\\
  \{\Gobs_{a_1},\Gobs_{a_2}\}_\bphspc &= \Gobs_{[a_1,a_2]}\period
\end{aligned}\end{equation}
Here $[a_1,a_2]$ is the Lie bracket of the vector fields $a_1$, $a_2$. In the
sense of the covariant phase space we have
\begin{equation}
  \{\Fobs_{f_1},\Fobs_{f_2}\}_\sphspc =
  \Fobs_{\grad f_1 \sint \gFC \sint \grad f_2}
  \quad\text{on}\quad \sphspc
\end{equation}%
  \index{Poisson brackets!on covariant phase space!\~ of basic observables $\Fobs_f$}%
and from the condition \eqref{SphspcInBphspc} we get
\begin{equation}
  \Gobs_a = \Fobs_{a\sint\grad\extr{S}}
  \quad\text{on}\quad \sphspc\period
\end{equation}

\subsection*{Summary}

Canonical phase spaces $\bphspc_\fix$, $\bphspc_\iix$ or the covariant phase
space $\sphspc$ are commonly used phase spaces of the classical theory. The
dynamical evolution is described as a canonical transformation of
$\bphspc_\iix$ to $\bphspc_\fix$ (the Schr\"odinger picture of the classical
theory) or as an evolution of observables on $\sphspc$ (Heisenberg picture on
the classical level).

If we could use experimental devices localized only on the boundary of the
investigated domain (i.e., if we could perform experiments only at the initial
and final moments of time), the boundary phase space $\bphspc$ would be
sufficient for a description of our system. It is enough to investigate
observables defined using canonical variables at the beginning and at the end.
The dynamics of the system is hidden in the definition of the special subspace
--- of the physical phase space $\sphspc$ that tells us the relation between
the initial value and momentum and the final value and momentum for a physical
solution of the equation of motion. In other words, the space $\bphspc$
represents all possible values of canonical observables at the initial and the
final time without knowledge of the equations of motion. The subspace $\sphspc$
represents values of canonical observables correlated via the dynamical
development of the system. The advantage of the boundary phase space $\bphspc$
is that we do not need any causal structure to define it. Hence, we can
construct it even for a Euclidian theory. This advantage will be even more
appealing in the quantum case.


\section{Configuration quantization}
\label{sc:Quant}

\subsection*{Overview}

In this chapter we describe a construction of quantum observables at most
linear in momenta for a general system with a phase space
$\phspc$. This chapter is completely independent of the previous one. The
starting point is a phase space with a cotangent bundle structure,  i.e., with
the phase space built over a general configuration space $\valspc$. We will
introduce a special type of quantum observables, that are generalizations of the
classical position and  momentum observables of the standard quantum mechanics
on a linear configuration space,  and we will find their position
representation.  The main idea is to understand the momentum observable as a
shift operator along some \vague{direction} in the configuration space. On a
general manifold, however, we have to be careful, because the
\vague{direction} is specified by a vector field that is, in general, position
dependent; and, of course, the position and momentum observables don't commute.

\subsection*{Ideas of quantization}

Quantization is a heuristic procedure of constructing a quantum theory for a
given classical theory. Let us have a classical system with a kinematical area
given by a phase space $\phspc$ and a symplectic structure $\gensymplstr$.
Observables are functions on $\phspc$, and the Poisson brackets are given by
Eq.~\eqref{PoisBrDef}. Quantization tells us how to assign to (at least some)
classical observables quantum observables --- operators on a quantum Hilbert
space $\quantspc$. We will use letters with a hat to denote quantum observables
and we denote $\qobsspc$ the algebra they obey. They should satisfy the same
algebraic relations as the classical observables and the commutation relations
generated by Poisson brackets. Specifically, if the quantum versions of
classical observables $A,B$, and $C = \{A,B\}$ are
$\qobs{A},\qobs{B},\qobs{C}$, they should be related by
\begin{equation}
  [\qobs{A},\qobs{B}] = -i \qobs{C}\commae
\end{equation}
$[\qobs{A},\qobs{B}] = \qobs{A}\qobs{B}-\qobs{B}\qobs{A}$ being the commutator
of operators. It is well known that the procedure described above cannot be
carried out for all classical observables. Because quantum observables do not
commute we have an \vague{ordering problem} for observables given by a product
of non-commuting observables.

The usual quantization procedure tries to quantize a specific class of
classical observables and construct physically interesting observables from
them. Even in this case the operator ordering ambiguity is encountered. But we
have to expect this --- a quantum theory is not fully determined by the
classical counterpart.

We will discuss quantization for a general example of a classical theory given
on a phase space $\phspc$ that has a cotangent bundle structure
$\tens^\dual\,\valspc$. The mathematical structure of such a phase space is
reviewed in Appendices \ref{apx:SySp} and \ref{apx:TSpG}. The main goal is not
to solve a dynamical problem here, but to formulate the quantization procedure
in a geometrically covariant way. We will define quantum observables of the
position and momentum in the case of a general configuration space $\valspc$
(i.e., without the usual assumption of linearity of the configuration space).
This procedure will be applied to quantization of the boundary phase space in
the next section.

Let us note that the procedure described below has a well-defined meaning for a
finite-dimensional configuration space $\valspc$. Of course, this is not the
case of, e.g., the scalar field theory, where the configuration space is the
space of functions on the space manifold~$\Sigma$. Technical problems with a
generalization to an infinite dimensional case is one of the reasons why this
kind of quantization is not usually employed to quantize a field theory, but
see \cite{Krtous:thesis} and the end of this section for an additional
discussion. However, in the following we will consider mainly a finite
dimensional non-relativistic system, for example a particle in a curved space.

\subsection*{Algebra of observables}

At the end of the last chapter we introduced two special kinds of observables
on the phase space with a cotangent bundle structure. We defined the
observables $\Fobs_f$ depending only on the \vague{position}
(Eq.~\eqref{FobsDef}) and the observables $\Gobs_a$ linear in momentum
(Eq.~\eqref{GobsDef}). Their Poisson brackets are given in
Eq.~\eqref{FGPoissBr}. Now, we will quantize these observables. First, we
formulate more carefully what conditions we are imposing on the quantum
versions of these observables.

We are looking for maps $\qFobs$ and $\qGobs$ from the test functions and test
vector fields on the configuration space $\valspc$ to the space of quantum
observables $\qobsspc$
\begin{equation}\label{FGobsNature}\begin{aligned}
  {}&\qFobs_f   \in \qobsspc &\quad&\text{for} \quad
    f\in\fctsct\,\valspc\commae\\
  {}&\qGobs_a   \in \qobsspc &\quad&\text{for} \quad
    a\in\tenssct\,\valspc\commae
\end{aligned}\end{equation}
that should be hermitian (for real $f$ and $a$)
\begin{equation}\label{FGobsHermiticity}
  \qFobs_f^\dagger = \qFobs_f \comma
  \qGobs_a^\dagger = \qGobs_a \commae
\end{equation}
and should satisfy commutation relations motivated by the Poisson brackets
\eqref{FGPoissBr}
\begin{align}
  [ \qFobs_{f_1},\qFobs_{f_2}] &= 0
    \commae\label{FFobsComRel}\\
  [ \qFobs_f,\qGobs_a] &= i\qFobs_{a\sint\grad f}
    \commae\label{FGobsComRel}\\
  [ \qGobs_{a_1},\qGobs_{a_2} ] &= -i \qGobs_{[a_1,a_2]}
    \period\label{GGobsComRel}
\end{align}

Next, we have to formulate the condition that the quantum observables satisfy
\vague{the same algebraic relations as the classical ones}. For the observables
$\qFobs_f$ this is straightforward because they commute --- we require that we
be able to take out any algebraic operation $g(\dots)$ from the argument of the
observable and perform the same operation on the observables:
\begin{equation}\label{FobsAlgDep}
  \qFobs_{g(f_1,f_2,\dots)} = g(\qFobs_{f_1},\qFobs_{f_2},\dots)
  \period
\end{equation}
With the $\qGobs_a$ we have to be more careful --- they do not commute with
each other and with the $\qFobs_f$ observables. But we are interested in
observables linear in momentum, so we need to investigate only the vector-field
dependence of the $\qGobs_a$. We require
\begin{gather}
  \qGobs_{a_1+\alpha a_2} = \qGobs_{a_1} + \alpha\qGobs_{a_2}
    \quad\text{for} \quad \alpha\in\realn
    \commae\label{GobsLinearity}\\
  \qGobs_{f a} = \qFobs_{f^{\frac12 - i \gamma}}
    \qGobs_a \qFobs_{f^{\frac12 + i \gamma}}
    \quad\text{for} \quad \gamma\in\realn
    \period\label{FGobsOrdering}
\end{gather}
The last condition is equivalent to the classical expression $\Gobs_{fa} =
\Fobs_f\Gobs_a$, but it specifies the ordering of the quantum theory. We have
chosen the exponents of the function $f$ in the special forms
$(\frac12-i\gamma)$ and $(\frac12+i\gamma)$ to satisfy the hermiticity
condition \eqref{FGobsHermiticity} with a real constant $\gamma$. The ordering
condition can be rewritten as
\begin{equation}
  \qGobs_{fa} =
  (\frac12-i\gamma)\, \qFobs_f\qGobs_a +
  (\frac12+i\gamma)\, \qGobs_a\qFobs_f =
  \frac12(\qFobs_f\qGobs_a + \qGobs_a\qFobs_f) +
  \gamma\,\qFobs_{a\sint\grad f}\period
\end{equation}

Finally, we require that the position observables $\qFobs_f$ form a complete set
of commuting observables
\begin{equation}\label{FobsCompl}
  \forall f \quad [\qFobs_f,\qobs{A}] = 0 \qquad \implr\qquad
  \exists\, g \quad \qobs{A} = \qFobs_g \period
\end{equation}
Let us note that from these conditions it follows that
\begin{gather}
  \qFobs_\alpha = \alpha \qone \quad\text{for}\quad
  \alpha\in\realn\commae\\
  \forall f,a \quad [\qFobs_f,\qobs{A}] = 0 \quad
  [\qGobs_a,\qobs{A}] = 0 \qquad \implr\qquad
  \exists\,\alpha\in\realn \quad \qobs{A} = \alpha\qone\period
\end{gather}

\subsection*{Position representation}

Next, we construct a \defterm{position base} in $\quantspc$ on which the action
of the operators $\qFobs_f$ and $\qGobs_a$ has a simple representation. Or, in
other words, we find a realization of the operators $\qFobs_f$ and $\qGobs_a$,
which satisfies the conditions formulated above, as operators on the space of
densities on the configuration manifold $\valspc$.

We have, of course, immediately a candidate for such a base. The observables
$\qFobs_f$ form a complete set of commuting observables, so there exists a base
of eigenvectors labeled by the position in the configuration space $\valspc$
such that
\begin{equation}\label{GenPosReprOfF}
  \qFobs_f |pos:\bhist{x}\rangle =
  f(\bhist{x})|pos:\bhist{x}\rangle\period
\end{equation}
Strictly speaking, $|pos:\bhist{x}\rangle$ are not vectors from the Hilbert
space $\quantspc$ but generalized vectors that can be defined, for example, by
the condition that the projectors to these states form an operator-valued
density on the configuration space. It will be convenient later to give these
vectors the character of a density of some (generally complex) weight
$\alpha\in\complexn$ (see Appendix~\ref{apx:Dens} for short review of densities
on a manifold). I.e., we assume
\begin{equation}
  |pos:\bhist{x}\rangle\in\quantspc\otimes
  \denst{\complexn}^\alpha_\bhist{x}\,\valspc\period
\end{equation}
The base of eigenvectors is orthogonal, but we need to be careful when writing
down a normalization condition. Because of the distributional character of the
vectors we can normalize them only to a delta-distribution. And to get the
right normalization we need to choose a volume element $\mu$ on the value
space. With such a volume element we can write the orthonormality relation
\begin{equation}\label{PosNormalization}
  \langle pos:\bhist{x}|pos:\bhist{y}\rangle =
  (\mu^{2\re\alpha-1}\deltadst)(\bhist{x}|\bhist{y}) \period
\end{equation}
The completeness relation is
\begin{equation}\label{PosCompletness}
  \qone = \int_\valspc |pos:.\rangle\langle pos:.|\:
  \mu^{1-2\re\alpha}\period
\end{equation}

Let us note that the conditions above do not fix the base
$|pos:\bhist{x}\rangle$ uniquely --- they fix the base up to an
$\bhist{x}$-dependent phase factor. We will deal with this ambiguity below.

For any vector $|state\rangle\in\quantspc$ we can define a \defterm{wave
function} --- a density of the weight $\alpha^\cxc$ on the configuration space
$\valspc$
\begin{equation}
  \wavefc_{|state\rangle}(\bhist{x}) = \langle pos:\bhist{x}|state\rangle\period
\end{equation}
The density $\mu$ defines the scalar product on this space which is an
isomorphic to the product on the quantum space
\begin{equation}
  \langle st 1 | st 2 \rangle =
  (\wavefc_{|st 1 \rangle},\wavefc_{|st 2 \rangle})_\mu =
  \int_\valspc \wavefc_{|st 1 \rangle}^\cxc
  \wavefc_{|st 2 \rangle}\, \mu^{1-2\re\alpha}\commae
\end{equation}
and it induces the hermitian conjugation ${}^\ddagger$ on operators acting on densities of
weight~$\alpha$
\begin{equation}\label{DensHSHermConj}
  ( A^\ddagger \psi_1,\psi_2 )_\mu = (\psi_1,A \psi_2)_\mu \period
\end{equation}
We define the \defterm{position representation} of a quantum observable
$\qobs{A}$ as an operator $\posrepr{A}$ on wave functions
\begin{equation}
  \posrepr{A}\, \wavefc_{|state\rangle} =
  \wavefc_{\qobs{A}|state \rangle}\period
\end{equation}
Clearly,
\begin{equation}\label{PosreprF}
  \posrepr{\Fobs}_f = f\deltadst \commae
\end{equation}
which can be also written as
\begin{equation}
  \qFobs_f = \int_\valspc f\, |pos:.\rangle\langle pos:.|\:
  \mu^{1-2\re\alpha}\period
\end{equation}

\subsection*{Phase fixing}

Now we proceed to find the position representation of the momentum
observables~$\qGobs_a$. In this section we show that there exists a unique choice of the
weight $\alpha$, namely $\alpha=(\frac12 -i\gamma)$, and of the phase of the
position base $|pos:\bhist{x}\rangle$, for which
\begin{equation}\label{PosreprG}
  \posrepr{\Gobs}_a = -i \liederr_a \commae
\end{equation}
where $\liederr_a$ is the Lie derivative along the vector field $a$ acting to
the right.

First we define the \defterm{position shift operator} along a vector field $a$
on the configuration space $\valspc$ as
\begin{equation}\label{PosShiftOp}
  \posshift_a(\varepsilon) = \exp(-i\varepsilon\qGobs_a)\period
\end{equation}
The commutation relation \eqref{FGobsComRel} gives us
\begin{equation}\label{PosshiftOnFobs}
  \posshift_a \qFobs_f \posshift_a^{\dash 1} =
  \qFobs_{\diffvf_a^\diffind f} \period
\end{equation}
Here $\diffvf_a(\varepsilon)$ is a diffeomorphism on $\valspc$ induced by the
vector field $a$
\begin{equation}
  \frac{d}{d\varepsilon}\diffvf_a = a \comma
  \diffvf_a(\varepsilon_1+\varepsilon_2) =
  \diffvf_a(\varepsilon_1)\,\diffvf_a(\varepsilon_2)\commae
\end{equation}
and $\diffvf_a^\diffind$ is a map induced by the diffeomorphism on objects
defined on the configuration space. The equation \eqref{PosshiftOnFobs} gives
us
\begin{equation}\begin{gathered}
  \qFobs_f\, \posshift_a |pos:\bhist{x}\rangle
  = \posshift_a \qFobs_{\diffvf_a^{\diffind\,\dash 1}f}
  |pos:\bhist{x}\rangle = f(\diffvf_a \bhist{x})\,
  \posshift_a |pos:\bhist{x}\rangle \qquad\implr\\
  \implr\qquad \posshift_a |pos:\bhist{x}\rangle \quad
  \text{is proportional to}\quad |pos:\diffvf_a \bhist{x}\rangle \period
\end{gathered}\end{equation}
Here we have to be careful about the proportionality coefficient because
vectors $|pos:\bhist{x}\rangle$ and $|pos:\diffvf_a \bhist{x} \rangle$ are also
densities in different points $\bhist{x}$ and $\diffvf_a \bhist{x}$. Because
$\posshift_a(\varepsilon)$ forms a commuting one-dimensional group for
$\varepsilon\in\realn$ we can write the proportionality relation as follows
\begin{equation}\label{UActionOnPsiPos}
  \diffvf_a^\diffind \Bigl(\posshift_a \Psi_a(\bhist{x})
  |pos:\bhist{x} \rangle\Bigr) =
  \Psi_a(\diffvf_a \bhist{x}) |pos:\diffvf_a \bhist{x} \rangle \commae
\end{equation}
where $\Psi_a$ is a density on $\valspc$ that is defined up to a density
invariant under the action of the diffeomorphism $\diffvf_a$.

Next we prove that $\Psi_a$ can be chosen as a density of weight
$(\frac12-\alpha)$ in the form
\begin{equation}\label{PsiFormLemmaOne}
  \Psi_a = \mu^{\frac12 - \alpha} \exp(-i\phi_a)\commae
\end{equation}
where $\phi_a$ is a real function on $\valspc$ defined up to a function
constant on the orbits of $\diffvf_a$.

We write $\Psi_a$ as a density
of the weight $(\frac12-\alpha)$ in the form
\begin{equation}
  \Psi_a = \mu^{\frac12 - \alpha} \rho_a \exp(-i\phi_a)\commae
\end{equation}
with $\rho_a$, $\phi_a$ real functions. Remembering Eq.~\eqref{PosShiftOp}, the
differential form of equation \eqref{UActionOnPsiPos}, gives\footnote{
  Let us recall that the Lie derivative of an object $A$ on the manifold is
  defined by
  \begin{equation*}
    (\lieder_a A)(\bhist{x}) =
    -\frac{d}{d\varepsilon}
    (\diffvf_a^\diffind A)(\bhist{x}) \Big|_{\varepsilon=0}=
    \frac{d}{d\varepsilon}
    \diffvf_a^{\diffind\,\dash 1} A(\diffvf_a\bhist{x})
    \Big|_{\varepsilon=0} \period
  \end{equation*}
  }
\begin{equation}
  \qGobs_a\Bigl(\mu^{\frac12 - \alpha}
  \rho_a \exp(-i\phi_a)|pos:. \rangle\Bigr) =
  i \lieder_a \Bigl(\mu^{\frac12 - \alpha}
  \rho_a \exp(-i\phi_a)|pos:. \rangle\Bigr) \period
\end{equation}
From this, it follows that the position representation of the $\qGobs_a$
observables:
\begin{equation}
  \posrepr{\Gobs}_a =
    -i\liederr_a
    - i \Bigl(
    (\frac12-\alpha^\cxc)\frac1{\mu}\lieder_a\mu +
    \frac1{\rho_a}\lieder_a\rho_a +
    i a\sint\grad\phi_a
    \Bigr)\deltadst\period
\end{equation}
The definition \eqref{DensHSHermConj} gives us
\begin{equation}
  \liederr_a^\ddagger = -\liederr_a +
  (2\re\alpha-1)\frac1\mu\bigl(\lieder_a\mu\bigr)\deltadst\period
\end{equation}
The hermiticity condition \eqref{FGobsHermiticity} implies $\posrepr{\Gobs}_a =
\posrepr{\Gobs}_a^\ddagger$. Substituting to this condition we obtain
\begin{equation}
  \frac1{\rho_a}\lieder_a\rho_a = 0 \commae
\end{equation}
which means $\rho_a$ should be constant, e.g., $\rho_a=1$. It proves the
statement \eqref{PsiFormLemmaOne}.

Finally we show that the function $\phi_a$ in equation \eqref{PsiFormLemmaOne}
has the form
\begin{equation}\label{PsiFormLemmaTwo}\begin{gathered}
  \phi_a = \varphi + \tilde\phi_a \commae\\
  \tilde\phi_a(\diffvf_a(\varepsilon) \bhist{x}) -
  \tilde\phi_a(\bhist{x}) =
  \gamma \int_0^\varepsilon
  \frac1{\mu} \bigl(\lieder_a\mu\bigr) \dvol{\varepsilon} \commae
\end{gathered}\end{equation}
where $\varphi$ is a real function independent of the vector field $a$.

Let us define a function
\begin{equation}\label{OmegaADef}
  \lambda_a = a\sint\grad\phi_a -
  \gamma \frac1{\mu} \lieder_a\mu\period
\end{equation}
This allows us to write $\posrepr{\Gobs}_a$ as
\begin{gather}
  \posrepr{\Gobs}_a = \posrepr{\Gobs}'_a + \lambda_a\deltadst\commae\\
  \posrepr{\Gobs}'_a =
    -i\liederr_a
    - i (\frac12+i\gamma-\alpha^\cxc)
    \frac1{\mu}\bigl(\lieder_a\mu\bigr) \deltadst
    \label{PosreprGAlpha}\period
\end{gather}
It is easy to check that the operators $\posrepr{\Gobs}'_a$ have the same
properties as the operators $\posrepr{\Gobs}_a$. Using the consequences of the
properties \eqref{GobsLinearity}, \eqref{FGobsOrdering}, and
\eqref{GGobsComRel}, we get conditions on $\lambda_a$:
\begin{equation}\begin{gathered}
  \lambda_{a_1+f a_2} = \lambda_{a_1} + f\lambda_{a_2}\commae\\
  \lambda_{[a_1,a_2]} = a\sint\grad\lambda_{a_2} -
  a_2 \sint\grad\lambda_{a_1}\period
\end{gathered}\end{equation}
The first condition implies that $\lambda_a = a\sint\lambda$ for some form
$\lambda$ on the configuration space $\valspc$ and the second condition implies
that this form is closed: $\grad\lambda = 0$. So, if we ignore topological
problems (see some of the comments below), we can rewrite relation
\eqref{OmegaADef} as
\begin{equation}
  a\sint\grad\phi_a = a\sint\grad\varphi +
  \gamma \frac1{\mu} \lieder_a\mu
\end{equation}
for a real function $\varphi$. Integrating along orbits of $\diffvf_a$, we get
the desired expression \eqref{PsiFormLemmaTwo}.

If we redefine our position base by the phase factor $\exp(-i\varphi)$, we
obtain the position representation for the momentum observables
$\posrepr{\Gobs}_a$ in the form \eqref{PosreprGAlpha}. We see that if we choose
the density weight of our position base to be
\begin{equation}
  \alpha=\frac12-i\gamma \commae
\end{equation}
the position representation reduces to the simple form \eqref{PosreprG}. This
also allows us to write the action of $\qGobs_a$ in the position base:
\begin{equation}\label{simpleGenPosReprOfG}
  \qGobs_a |pos:. \rangle = i \lieder_a |pos:. \rangle \period
\end{equation}

Let us note that for this choice of $\alpha$ we do not need any volume element
on $\valspc$ because the normalization and completeness conditions
\eqref{PosNormalization} and \eqref{PosCompletness} reduce to
\begin{equation}\label{PosNormComplSpec}
  \langle pos:\bhist{x}|pos:\bhist{y}\rangle =
  \deltadst(\bhist{x}|\bhist{y}) \comma
  \qone = \int_\valspc |pos:.\rangle\langle pos:.|\period
\end{equation}

\subsection*{Uniqueness of the quantization}

Now we ask the question whether all realizations of our quantization of the
basic observables $\Fobs_f$ and $\Gobs_a$ are unitarily equivalent. Let us
assume we have two quantum versions of the basic observables $\qFobs_f$,
$\qGobs_a$ and $\qFobs'_f$, $\qGobs'_a$ that both satisfy all the conditions
formulated above. Clearly, we can construct a unitary operator mapping the
position base of the first pair to the position base of the second pair, except
that we do not require a proper phase fixing. Such a unitary operator
essentially identifies the position observables but not necessarily the
momentum observables.

Therefore, we will investigate the relation between the observables $\qGobs_a$
and $\qGobs'_a$ that both, together with the common position observables
$\qFobs_f$, satisfy all the conditions above. The commutation relation
\eqref{FGobsComRel} together with the completeness condition \eqref{FobsCompl}
gives
\begin{equation}
  \qGobs'_a - \qGobs_a = \qFobs_{\lambda_a}
\end{equation}
for an $a$-dependent real function $\lambda_a$. The linearity
\eqref{GobsLinearity} and the ordering condition \eqref{FGobsOrdering} together
with the property \eqref{FobsAlgDep} give
\begin{equation}
  \lambda_{a_1 + f a_2} = \lambda_{a_1} + f \lambda_{a_2}
  \qquad\implr\qquad \lambda_{a} = \lambda\sint a
\end{equation}
for some form $\lambda$ on $\valspc$. The commutation relation
\eqref{GGobsComRel} implies
\begin{equation}
  [a_1,a_2] \sint \lambda =
  a_1\sint\grad(a_2\sint\lambda)  -
  a_2\sint\grad(a_1\sint\lambda)
  \qquad\implr\qquad \grad\lambda = 0\period
\end{equation}

If the configuration space $\valspc$ is sufficiently topologically trivial
(precisely, if the first cohomology group is trivial), it follows from the last
equation that the form $\lambda$ is a gradient of some function $\varphi$. In
this case we can write
\begin{equation}
  \qGobs'_a = \exp(-i\qFobs_\varphi) \qGobs_a \exp(i\qFobs_\varphi)\commae
\end{equation}
and we see that $\qGobs'_a$ and $\qGobs_a$ are unitary equivalent. If the
configuration space is not topologically trivial and closed forms are not the
same as exact forms, we can have unitarily inequivalent realizations of the
basic quantum variables. We ignored this possibility when constructing the
position base and we will not investigate it further here.

\subsection*{Relation between different orderings}

Here we will investigate a relation between different orderings of the momentum
observables. Let us assume that ${}^{\gamma_1}\qGobs$ and ${}^{\gamma_2}\qGobs$
together with the position observables $\qFobs_f$ satisfy the above conditions
with the parameter $\gamma_1$ or $\gamma_2$ in the ordering condition
\eqref{FGobsOrdering}. Similary to the previous discussion we get
\begin{gather}
  {}^{\gamma_1}\qGobs_a - {}^{\gamma_2}\qGobs_a =
  \qFobs_{\tilde\lambda_a}\commae\\
  \tilde\lambda_{a_1 + \alpha a_2} =
  \tilde\lambda_{a_1} + \alpha \tilde\lambda_{a_2}
  \quad\text{for}\quad \alpha\in\realn\commae\\
  \tilde\lambda_{f a} =  f \tilde\lambda_{a} +
  (\gamma_1 - \gamma_2) a\sint\grad f \period
\end{gather}
If we write the function $\tilde\lambda_a$ using a density $\mu$ on $\valspc$
as
\begin{equation}
  \tilde\lambda_{a} =
  (\gamma_1-\gamma_2) \frac1\mu \lieder_a \mu
  + \lambda_a\commae
\end{equation}
we find that $\lambda_a$ has the same properties as in the previous section,
i.e., it represents the freedom of the quantization of $\Gobs_a$ with the given
ordering parameter. Because we have discussed it already, we ignore it now.
Thus, we found that different orderings of the momentum observable can be
written in the form
\begin{equation}
  {}^{\gamma}\qGobs_a = {}^0\qGobs_a + \gamma\,
  \qFobs_{\mu^{\dash1} \lieder_a \mu} \period
\end{equation}
It is easy to check that the $\mu$ dependence for a fixed $\gamma$ is of the
form discussed in the previous section.

\subsection*{Observables quadratic in momentum}

Until now we only discussed quantization of observables independent of momentum
and linear in momentum. We saw that these observables were sufficient for the
construction of the natural base in the quantum Hilbert space $\quantspc$. But
they are not usually sufficient for the construction of the dynamics of the
theory. A typical Hamiltonian is quadratic in momenta. Therefore, it is
necessary to address the issue of quantization of such observables. I.e., we
want to quantize classical observables of the form
\begin{equation}
  \Kobs_k(\bhist{x},\bhist{p}) =
  \bhist{p}\sint k^{\dash 1}(\bhist{x}) \sint \bhist{p}\commae
\end{equation}
where $k$ is a metric on $\valspc$ and we have restricted ourself to the case
of a non-degenerate $k$.

We can formulate conditions similar to those above for $\qFobs_f$ and
$\qGobs_a$. With a suitable choice of simplicity and covariance requirements it
is possible to show that there is a one-dimensional freedom in the ordering for
the quadratic observables (labeled by a parameter $\xi\in\realn$) and that the
position representation $\posrepr{\Kobs}_k$ of the quadratic observable is
\begin{equation}
  \posrepr{\Kobs}_k = \scdst{L}_k + \xi \mathcal{R}_k \commae
\end{equation}
where $\scdst{L}_k$ and $\mathcal{R}_k$ are the Laplace operator and scalar
curvature of the metric $k$.

\subsection*{Linear theory}

The situation simplifies significantly if the configuration space $\valspc$ is
linear. This is usually assumed in simple quantum mechanical models and it is
essential in the quantum field theory. For a linear theory, the formalism
reduces to the standard canonical quantization. We will shortly illustrate the
correspondence with the usual formalism using a notation borrowed from the the
scalar field theory. However, these notes apply for any linear theory.

In the scalar field theory, the configuration space $\valspc[\Sigma]$
represents the space of all field configurations on a Cauchy hypersurface
$\Sigma$ and the phase space $\bphspc[\Sigma]$ has the structure of a cotangent
bundle $\tens^\dual\,\valspc[\Sigma]$ that can be identified with the dual
space $\valspc[\Sigma]^\dual$. The dual can be realized as the space of
densities $\momspc[\Sigma]$ of the weight $1$ on the hypersurface $\Sigma$.
This brings good and bad news. As we said, the value space $\valspc[\Sigma]$ is
linear, but it is infinite-dimensional. Let's ignore the infinite dimension and
do some formal manipulation first.

Linearity allows us to define the observables of value and momentum $\qvalobs$
and $\qmomobs$ (roughly speaking operators $\qobs{\bhist{x}}$ and
$\qobs{\bhist{p}}$ in the previous notations). These are, of course, not well
defined objects in the general non-linear case but in the case of a linear
space $\valspc$ it is possible to define them as objects from the spaces
$\valspc\otimes\qobsspc$ and $\momspc\otimes\qobsspc$. They are connected with
the general observables $\qFobs_f$, $\qGobs_a$ of the previous sections as
\begin{equation}\begin{gathered}
  \qFobs_f = f(\qvalobs) \commae\\
  \qGobs_a = (\frac12-i\gamma)\: a(\qvalobs)\sint\qmomobs \,+\,
  (\frac12+i\gamma)\: \qmomobs\sint a(\qvalobs)\commae
\end{gathered}\end{equation}
where we used the natural identification of the tangent space $\tens\,\valspc$
with $\valspc$ itself. The value and momentum observables satisfy the canonical
commutation relations
\begin{equation}\label{ValMomComRel}
  [\qvalobs,\qmomobs] = i \deltadst_\valspc  \qone\period
\end{equation}

We can construct the \defterm{value base} normalized to a \vague{constant
measure} $\measure{Q}$ on the configuration space $\valspc$
\begin{equation}\label{ValBase}
\begin{gathered}
  \qvalobs |val:\varphi \rangle  = \varphi |val:\varphi \rangle
  \quad\text{for}\quad \varphi\in\valspc\commae\\
  \qmomobs |val:\varphi \rangle  = i\: \grad |val:\varphi \rangle\commae\\
  \langle val: \varphi_1|val:\varphi_2 \rangle =
  (\measure{Q}^{\dash 1}\deltadst)(\varphi_1|\varphi_2)\commae\\
  \qone = \int_{\varphi\in\valspc} \measure{Q}\,
  |val: \varphi \rangle \langle val:\varphi|\period
\end{gathered}\end{equation}
The wave functional for a state $|state \rangle$ has the form
\begin{equation}
  \wavefc_{|state \rangle}(\varphi) =
  \langle val:\varphi|state \rangle\commae
\end{equation}
the scalar product on wave functions is
\begin{equation}\label{WaveScPr}
  \langle st 1 | st 2 \rangle =
  \int_\valspc \wavefc_{| st1 \rangle}^\cxc
  \wavefc_{| st 2 \rangle}\, \measure{Q} \commae
\end{equation}
and the value representations of the observables $\qvalobs$ and $\qmomobs$ are
\begin{equation}\begin{gathered}
  \posrepr{\valobs}\,\wavefc_{|state \rangle}(\varphi) =
  \varphi\,\wavefc_{|state \rangle}(\varphi)\commae\\
  \posrepr{\momobs}\,\wavefc_{|state \rangle}(\varphi) =
  -i\, \grad\wavefc_{|state \rangle}(\varphi)\period
\end{gathered}\end{equation}

The problem is that in the infinite-dimensional case there is no such thing as
a \vague{constant measure} $\measure{Q}$ on the space $\valspc$ and we do not
have a Hilbert space generated by a scalar product on wave functions. The
solution to these technical difficulties lies in a restriction of the possible
wave functions to those that are falling off sufficiently fast so the
functional integral \eqref{WaveScPr} has a meaning even if the measure
$\measure{Q}$ itself does not have one. Such wave functions have to be
suppressed, for example, by a Gaussian exponent
--- so the integral \eqref{WaveScPr} turns into a Gaussian integration that is
well defined even for infinite-dimensional spaces.


\section{Boundary Quantum Mechanics}
\label{sc:BQMGen}

\subsection*{Introduction}

In this chapter we will finally develop \defterm{boundary quantum mechanics}
--- a variation of quantum mechanics based on quantization of the boundary
phase space. Here we present a general formulation of the method, an
application to scalar field theory can be found in \cite{Krtous:thesis}.

Let us start with the quantum theory described in the previous section. We
quantized the observables $\Fobs_f$ and $\Gobs_a$ on a phase space with the
cotangent bundle structure, we found quantum operators $\qFobs_f$ and
$\qGobs_a$, and we constructed the special position base $|pos: \bhist{x}
\rangle$. We did not formulate the dynamical part of the theory; we will touch
this issue now.

We are interested in a situation when we study the system only at the
\vague{beginning} and at the \vague{end}. More precisely, we are interested in
observables with a support only on the boundary of the domain $\Omega$. It
means that in the case of the field theory, when the inner manifold is a
globally hyperbolic spacetime, observables are localized only on the initial
and final hypersurfaces of the sandwich domain $\Omega = \langle \Sigma_\fix,
\Sigma_\iix \rangle$. In case of a one-dimensional inner space manifold, the
boundary of $\Omega$ reduces to the initial and final moment of time. In these
cases we can quantize the initial and final canonical phase spaces
$\bphspc_\iix=\bphspc[\Sigma_\iix]$ and $\bphspc_\fix=\bphspc[\Sigma_\fix]$ ---
i.e. we construct the quantum observables $\qFobs_{\fix\,f}$,
$\qGobs_{\fix\,a}$ and $\qFobs_{\iix\,f}$, $\qGobs_{\iix\,a}$ on a quantum
space $\quantspc$ satisfying the conditions
(\ref{FGobsNature}--\ref{FobsCompl}) with the ordering parameters $\gamma_\fix
= - \gamma_\iix$, and the corresponding position bases\footnote{
  We use the proper normalization of the position bases and choose the
  density weight of the base equal to $(\frac12 - i \gamma_\fix)$ and
  $(\frac12 - i\gamma_\iix)$, where $\gamma_\fix$ and $\gamma_\iix$ are the
  order parameters as defined in \eqref{FGobsOrdering}. I.e. we do
  not need to choose any volume element on $\valspc[\Sigma_\fix]$ or
  $\valspc[\Sigma_\iix]$ and it makes sense to write, for example,
  \[
    \qFobs_{\fix\,f} = \int_{\bhist{x}_\fix\in\valspc[\Sigma_\fix]}
    |\fix\,pos: \bhist{x}_\fix \rangle
    \langle \fix\,pos: \bhist{x}_\fix |\period
  \]}
$|\fix\,pos: \bhist{x}_\fix \rangle$ and $|\iix\,pos: \bhist{x}_\iix \rangle$
in the space $\quantspc$. The dynamics reduces to the investigation of
relations between these two sets of observables or relations between objects
generated by them, for example, of the position bases. We can state we solved
the dynamical problem if we find in-out transition amplitudes $ \langle
\fix\,pos: \bhist{x}_\fix |\iix\,pos: \bhist{x}_\iix \rangle$ for all $
\bhist{x}_\fix\in\valspc[\Sigma_\fix]$ and
$\bhist{x}_\iix\in\valspc[\Sigma_\iix]$.

We do not attempt to find the transition amplitudes in a general situation.
Instead, we reformulate this setting in a slightly different language.

\subsection*{Construction of the boundary quantum space}

We represented the quantization of both $\bphspc_\fix$ and $\bphspc_\iix$ on a
common quantum Hilbert space $\quantspc$. Let us construct another
representation on the \defterm{boundary quantum space}
\begin{equation}
  \quantspc_\bphspc = \quantspc^\dagger\otimes\quantspc
\end{equation}
(the space of tensor products of covector and vector elements of the form $
\langle \fix | \otimes | \iix \rangle$, i.e., essentially operators on
$\quantspc$). We use the notation $|state)$ for vectors from
$\quantspc_\bphspc$, and we use the accent $\Qobs{\rule{0cm}{1.4ex}}$ to denote
observables on this space.

We interpret the boundary quantum space in the following way: we assign a
vector $|\fix\iix)$ from $\quantspc_\bphspc$ to any pair of $\fix$-dependent
and $\iix$-dependent vectors $|\fix \rangle$ and $|\iix \rangle$ by
\begin{equation}
  |\fix\iix) = \langle \fix | \otimes |\iix \rangle \period
\end{equation}
Here, the description \emph{$\fix$-dependent vector} suggests that the vector
is identified using quantum observables on $\Sigma_\fix$; but, of course, it
can be any vector from $\quantspc$. Particularly, we define a vector in
$\quantspc_\bphspc$ for any $\bhist{x} =
[\bhist{x}_\fix,\bhist{x}_\iix]\in\valspc[\bound\Omega]=
\valspc[\Sigma_\fix]\times\valspc[\Sigma_\iix]$:
\begin{equation}
  |pos:\bhist{x} ) = \langle \fix\, pos: x_\fix |
  \otimes | \iix\, pos: \bhist{x}_\iix \rangle \period
\end{equation}

This allows us to \vague{lift} any $\fix$-dependent or $\iix$-dependent
operator $\qobs{A}_\fix$ or $\qobs{A}_\iix$ on $\quantspc$ to an operator on
the boundary quantum space $\quantspc_\bphspc$
\begin{equation}\begin{aligned}
  \Qobs{A}_\fix &= \qobs{A}_\fix^\dagger\otimes\qone_\iix &\qquad
  \Qobs{A}_\fix |\fix\iix ) &=
  \langle\fix |\qobs{A}_\fix^\dagger\otimes |\iix \rangle \commae\\
  \Qobs{A}_\iix &= \qone_\fix\otimes\qobs{A}_\fix &\qquad
  \Qobs{A}_\iix |\fix\iix ) &=
  \langle\fix |\otimes\qobs{A}_\iix |\iix \rangle \period
\end{aligned}\end{equation}
Using these definitions, we find
\begin{equation}\label{ifObsCommute}
  [\Qobs{A}_\fix,\Qobs{A}_\iix] = 0 \period
\end{equation}

We apply this method to construct the observables $\QFobs_{\fix\,f_\fix}$,
$\QGobs_{\fix\,a_\fix}$ and $\QFobs_{\iix\,f_\iix}$, $\QGobs_{\iix\,a_\iix}$
for any functions and vector fields $f_\fix$, $a_\fix$ or $f_\iix$, $a_\iix$ on
the value spaces $\valspc[\Sigma_\fix]$ or $\valspc[\Sigma_\iix]$. Next we want
to construct a generalization of these observables $\QFobs_f$ and $\QGobs_a$
for any function $f$ and vector field $a$ on the value space
$\valspc[\bound\Omega] = \valspc[\Sigma_\fix]\times\valspc[\Sigma_\iix]$.
Thanks to \eqref{FFobsComRel} for both $\qFobs_{\fix\,f}$ and
$\qFobs_{\iix\,f}$ and to equation \eqref{ifObsCommute}, we do not have
ordering problems with $\QFobs_f$ for $f(\bhist{x}) =
f(\bhist{x}_\fix,\bhist{x}_\iix)$
\begin{equation}
  \begin{split}
  \QFobs_f &= \int_{\substack{\bhist{x}_\fix\in\valspc[\Sigma_\fix]\\
  \bhist{x}_\iix\in\valspc[\Sigma_\iix]}}
  |\fix\,pos: \bhist{x}_\fix \rangle
  \langle\fix\, pos: \bhist{x}_\fix |
  \otimes |\iix\,pos: \bhist{x}_\iix \rangle
  \langle\iix\, pos: \bhist{x}_\iix |\;
  f(\bhist{x}_\fix, \bhist{x}_\iix) =\\
  &= \int_{\bhist{x}\in\valspc[\bound\Omega]}
  |pos: \bhist{x} )(pos: \bhist{x}|\, f(\bhist{x}) \period
\end{split}
\end{equation}
Similarly, for any vector field $a(\bhist{x})=a_\fix(\bhist{x}_\fix)\oplus
a_\iix(\bhist{x}_\iix)$ on $\valspc[\bound\Omega]$ where
$a_\fix\in\tenssct\,\valspc[\Sigma_\fix]$ and
$a_\iix\in\tenssct\,\valspc[\Sigma_\iix]$, motivated by \eqref{fidecompofp}, we
can write
\begin{equation}
  \QGobs_a = - \QGobs_{\fix\,a_\fix} + \QGobs_{\iix\,a_\iix}\period
\end{equation}

It is straightforward to check that $\QFobs_f$, $\QGobs_a$ are quantizations of
the observables $\Fobs_f$, $\Gobs_a$ on the boundary phase space
$\bphspc[\bound\Omega]$; i.e. they satisfy (\ref{FGobsNature}-\ref{FobsCompl})
with the ordering parameter $\gamma=\gamma_\iix =\gamma_\fix$. Note that a
different orientation of the boundary $\bound\Omega$ and the final hypersurface
$\Sigma_\fix$, which translates to the different sign of the symplectic
structures \eqref{fisymplstrdecomp} and to the definition of the momenta
\eqref{fidecompofp}, is compensated by the covector representation of
$\fix$-dependent vectors.

Let us summarize: quantization of the basic observables on the final and
initial hypersurfaces $\Sigma_\fix$ and $\Sigma_\iix$ induces quantization of
the basic observables $\Fobs_f$, $\Gobs_a$ on the entire boundary
$\bound\Omega$. Hence, we are able to formulate the \vague{kinematics} of the
theory using quantization of the boundary phase space $\bphspc[\bound\Omega]$
--- which we call \defterm{boundary quantum mechanics}. The boundary
quantum space $\quantspc_\bphspc$ represents all possible quantum states at the
beginning and at the end chosen independently of the real evolution of the
system. Essentially, we are treating the initial and final experiments as
experiments on independent systems. States in $\quantspc_\bphspc$ represent
outputs of measurements understood in this way.

Before we turn to the dynamics let us list some properties of the space
$\quantspc_\bphspc = \quantspc^\dagger \otimes \quantspc$. For vectors and
operators in the \vague{product} form
\begin{equation}\label{ProdForm}\begin{gathered}
  |\fix\iix) = \langle \fix | \otimes |\iix \rangle\commae\\
  \Qobs{O}_{\fix\iix} = \qobs{O}_\fix^\dagger \otimes \qobs{O}_\iix\commae
\end{gathered}\end{equation}
we can write
\begin{equation}\begin{gathered}
  (\fix\iix\,st1|\fix\iix\,st2) =
  \langle\fix\,st2|\fix\,st1 \rangle\,
  \langle\iix\,st1|\iix\,st2 \rangle =
  \Tr_\quantspc \bigl((|\fix\,st2 \rangle \langle\fix\,st1|)^\dagger\;
  (|\iix\,st2 \rangle \langle\iix\,st1 |)\bigr)\commae\\
  (\fix\iix| = |\fix\iix)^\dagger =
  |\fix\rangle\otimes \langle\iix| \comma
  \Qobs{A}_{\fix\iix}\Qobs{B}_{\fix\iix} =
  (\qobs{B}_\fix^\dagger\qobs{A}_\fix^\dagger) \otimes
  (\qobs{B}_\iix\qobs{A}_\iix)\commae\\
  \Qobs{O}_{\fix\iix}|\fix\iix) =
  \langle \fix|\qobs{O}_\fix^\dagger\otimes
  \qobs{O}_\iix |\iix \rangle \comma
  \Tr_{\quantspc_\bphspc} \Qobs{O}_{\fix\iix} =
  \Tr_\quantspc \qobs{O}_\fix^\dagger \;
  \Tr_\quantspc \qobs{O}_\iix \period
\end{gathered}\end{equation}

\subsection*{Dynamics in the boundary quantum space}

Of course, much more interesting is the dynamical part of a theory. We have to
ask the question whether we are able to translate the dynamically interesting
quantities to the language of boundary quantum mechanics. As we outlined, the
dynamical information is hidden in the in-out transition amplitudes $
\langle\fix | \iix \rangle$. Such an amplitude can be written as
\begin{equation}
  \langle \fix | \iix \rangle =
  \Tr_\quantspc \bigl(\qone^\dagger \, |\iix \rangle \langle \fix | \bigr) =
  (phys|\fix\iix) \commae
\end{equation}
where $|\fix\iix)$ is as in \eqref{ProdForm} and the \defterm{physical state
$|phys)$} is given by
\begin{equation}
  |phys) = \sum_{\mathrm k}
  \langle {\mathrm k} | \otimes | {\mathrm k} \rangle = \qone
\end{equation}
for a complete orthonormal base $|{\mathrm k} \rangle$ in $\quantspc$.

This means that there exists a preferred physical state $|phys)$ in the
boundary quantum space $\quantspc_\bphspc$ that determines the dynamics of the
theory. Specifically, if we set up some initial and final experiments that
determine the quantum state $|state)\in\quantspc_\bphspc$, the physical
transition amplitude corresponding to this state is given by
\begin{equation}
  \ampl{A}(state) = (phys|state) \period
\end{equation}
For example, for the position base $|pos: \bhist{x})$ we get
\begin{equation}
  \ampl{A}_{pos}(\bhist{x}) = \ampl{A}_{pos}(\bhist{x}_\fix|\bhist{x}_\iix) =
  (phys|pos: \bhist{x}) =
  \langle\fix\,pos:\bhist{x}_\fix|\iix\,pos:\bhist{x}_\iix \rangle\period
\end{equation}
We will call $\ampl{A}_{pos}(\bhist{x})$ the \defterm{position transition
amplitude}.

The physical state $|phys)$ is actually an entangled quantum state that carries
all information about the dynamical correlations between the initial and final
moment of time without reference to particular initial or final conditions.

\subsection*{Boundary quantum mechanics}

In the previous subsections we constructed the boundary quantum space and
observables on it using quantization based on the initial and final phase
spaces. But it is clear that we can skip the splitting of the boundary into two
pieces and quantize directly the basic observables $\Fobs_f$, $\Gobs_a$ on the
boundary phase space $\bphspc[\bound\Omega]$. It is a phase space with a
cotangent bundle structure, so we can apply the general formalism and obtain
the quantum observables $\QFobs_f$ and $\QGobs_a$. We can also construct the
position base $|pos: \bhist{x})$. And we do not need any causal information for
this; we do not need any global time flow on the underlying inner manifold or a
causal decomposition of the boundary.

It means that we can build boundary quantum mechanics even in situations where
we do not have any natural splitting of the boundary into two pieces, for
example in Euclidian theories. Therefore, we could call boundary quantum
mechanics also \defterm{time-symmetric quantum mechanics}.

However, in this setting we have to find the physical state $|phys)$ without
reference to the initial and final causal decomposition.\footnote{
  More precisely, in the Heisenberg picture, which we are using, the physical
  state $|phys)$ is a fixed dynamically independent state and the dynamics is
  hidden in the relations of the basic observables $\QFobs_f$, $\QGobs_a$ to this
  state. But we will be a bit vague and speak about a determination of the
  physical state $|phys)$ because it is more intuitive and does not influence
  any computation.}
Again, we cannot expect an answer on a general level --- this is a question
equivalent to solving the quantum evolution. But we suggest methods determining
the physical state --- we formulate a dynamical equation for boundary quantum
mechanics.

The classical evolution in the boundary phase space $\bphspc[\bound\Omega]$ is
determined by a specification of the physical phase space $\sphspc$ as a
subspace of $\bphspc[\bound\Omega]$. It can be done, for example, via condition
\eqref{SphspcInBphspc}, or, expressed using observables $\Fobs_f$, $\Gobs_a$,
by the condition
\begin{equation}
  \Gobs_a + \Fobs_{a\sint\grad\extr{S}} = 0 \qquad
  \text{for all} \quad a\in\tenssct\,\valspc[\bound\Omega]\period
\end{equation}
Hence, on the classical level we specified physical states by imposing
constraints in the phase space. In the usual quantum mechanics one quantizes
the constrained subspace $\sphspc$. In boundary quantum mechanics we quantize
the entire boundary phase space, but we have to impose conditions on the
physical state inspired by the classical constraints
\begin{equation}\label{PhysStConstr}
  \bigl(\QGobs_a + \QFobs_{a\sint\grad\extr{S}}\bigr) |phys) = 0
\end{equation}
for, at least, some vector fields $a$ on the value space $\valspc[\bound\Omega]$.

We cannot expect the condition above to be satisfied for all vector fields $a$
due to the noncommutativity of the position and momentum observables
($\QGobs_a$ is some ordering of the \vague{$a(\genval)\sint \genmom$}
observable). We will see that such a strong requirement would be inconsistent.
It will turn out that the choice of the class of vector fields for which the
condition \eqref{PhysStConstr} is required to hold (i.e. the choice of a
preferred operator ordering) is equivalent to the solution of the dynamical
problem. So, if we have some preferred vector fields, it can provide us with a
method for finding the transition amplitudes we are looking for.

Let us look at the position representation of the constraint conditions. Using
\eqref{GenPosReprOfF} and \eqref{simpleGenPosReprOfG} we find
\begin{equation}
  (phys|\bigl(\QGobs_a + \QFobs_{a\sint\grad\extr{S}}\bigr)|pos:.) =
  \bigl(i\lieder_a + a\sint\grad\extr{S}\bigr) (phys|pos: . ) = 0 \period
\end{equation}
If we represent the position transition amplitude as
\begin{equation}\label{AmplUsingMeasure}
  \ampl{A}_{pos}(\bhist{x}) =
  \measure{a}(\bhist{x})\,\exp\bigl(i \extr{S}(\bhist{x})\bigr)
\end{equation}
with $\measure{a}(\bhist{x})$ a density of the weight $(\frac12 - i\gamma)$,
the condition above translates to
\begin{equation}
  \lieder_a \measure{a} = 0 \period
\end{equation}
This confirms that the constraint conditions cannot be satisfied for all vector
fields~$a$ --- it would require ${\measure{a}=0}$.

If a linearly complete set of vector fields for which the constraint conditions
should be satisfied is specified, the density $\measure{a}$ is determined
completely up to a constant multiplicative factor (this remaining freedom of
choice is, of course, equivalent to the choice of a global normalization and
phase factor). In case of a linear non-interacting theory,\footnote{
  Motivated by the field theory, by \emph{non-interacting} theory we mean a
  Hamiltonian quadratic in momenta and positions.
  }
such a set of vectors exists --- if we require the condition
\eqref{PhysStConstr} be satisfied for globally parallel vector fields\footnote{
  The notion of global parallelism is defined thanks to the linearity.
  }
we obtain that $\measure{a}$ is a constant and we recover the known fact that
the quantum amplitudes for a non-interacting theory are given by the classical
action through \eqref{AmplUsingMeasure}.

In a general case, the density $\measure{a}$ contains all quantum corrections
to the transition amplitude $\ampl{A}_{pos}$. It indicates that a choice of the
\vague{right} vector fields in the case of a general interacting theory can be
difficult or even impossible. Therefore, we turn to another way of specifying
the physical state $|phys)$ or physical amplitudes $\ampl{A}(state)$.

\subsection*{Path integral}

There exists another approach to the quantization. This is the path integral
quantization, which gives essentially a prescription for the position
transition amplitude on the basis of a completely different calculation ---
through a sum of elementary amplitudes over all possible histories with fixed
boundary values,
\begin{equation}
  \langle\fix\,pos:\bhist{x}_\fix|\iix\,pos:\bhist{x}_\iix \rangle =
  \ampl{A}_{pos}(\bhist{x}_\fix|\bhist{x}_\iix) =
  \int_{\substack{\hist{h}\in\histset\\
  \genval(\hist{h}) = [\bhist{x}_\fix,\bhist{x}_\iix]}}
  \fmsr(\hist{h}) \:\exp\bigl(i S(\hist{h})\bigr) \period
\end{equation}
However, this amplitude is exactly the wave function of the physical state
$|phys)$, i.e., the dynamics of boundary quantum mechanics can be specified by
the path integral
\begin{equation}\label{LorPIOMRel}
  (phys| =
  \int_{\hist{h}\in\histset}
  \fmsr(\hist{h}) \:\exp\bigl(i S(\hist{h})\bigr)\;
  (pos:\genval(\hist{h})|\period
\end{equation}

Let us add some comments about the advantages and problems of this approach.
The integral over the space of histories faces serious problems due to the
infinite dimension of the space of histories and the oscillatory character of
the integrand. A technical solution usually leads to the computation of some
Euclidian equivalent of the integral, which is usually better defined, followed
by some \vague{Wick rotation} --- a transformation from the Euclidian to the
physical theory. We can view this \vague{Euclidian business} as a mere
technical detour without a physical interpretation. Only after computing the
integral we are able to identify the results with the transition amplitudes of
quantum mechanics. In the usual framework, we do not even know what would be
the quantum mechanics of the Euclidian formulation of the theory --- the usual
quantum mechanics essentially uses the causal structure to define the initial
and final states.

However, the formalism developed above gives us a hope for another option. We
can formulate boundary quantum mechanics even for a Euclidian version of the
theory --- it does not need the causal structure, it can be formulated without
splitting of the boundary phase space to the initial and final part. Therefore
we could make the connection with the path integral already in the Euclidian
formulation
\begin{equation}
  (phys|pos: \bhist{x}) = \ampl{A}_{pos}(\bhist{x}) =
  \int_{\substack{\hist{h}\in\histset\\
  \genval(\hist{h}) = \bhist{x}}}
  \fmsr(\hist{h}) \:\exp\bigl( - I(\hist{h})\bigr) \commae
\end{equation}
$I(\hist{h})$ being the Euclidian action.\footnote{
  $I(\hist{h})$ is real for \vague{Euclidian} histories.
  For physical histories we have $i S(\hist{h})=-I(\hist{h})$.
  }
For the physical version of the theory, this reduces to the relations
\eqref{LorPIOMRel} above. In the Euclidian case, this relation would give an
interpretation for the path integral amplitude in terms of transition
amplitudes of boundary quantum mechanics.

Unfortunately, the situation is not so straightforward. Even in case of a
linear non-interacting theory, the form of boundary quantum mechanics
formulated above applied to the Euclidian version of the theory does not
correspond exactly to the Euclidian path integral. The problem is hidden in the
method of quantization we used. The translation of the Poisson brackets of the
classical theory to the commutators of the quantum theory
\begin{equation}
  \{\;,\,\} \rightarrow i[\;,\,]
\end{equation}
intrinsically contains reference to the physical signature. The imaginary unit
in this translation can be traced to be the cause of the imaginary unit in the
exponent of the transition amplitude \eqref{AmplUsingMeasure} computed in
boundary quantum mechanics independently of the physical or Euclidian version
of the theory. For the Euclidian theory it would be more appropriate to
construct \vague{Euclidian quantum mechanics} based on the commutation relation
generated by the rule
\begin{equation}\label{EucliQuantRule}
  \{\;,\,\} \rightarrow [\;,\,] \commae
\end{equation}
with observables represented, probably, on a real Hilbert space. We do not
attempt to build such quantum mechanics here. However, escaping the necessity
of a causal structure in the usual quantum mechanics through the method of
quantization of the boundary phase space is the first step towards Euclidian
quantum mechanics.

\appendices


\section{Symplectic geometry}
\label{apx:SySp}


This appendix is a review of the standard geometrical formulation of the
symplectic geometry (see, e.g., \cite{Arnold:book,Frankel:book}).

The phase space is a manifold $\phspc$ of an even dimension $2 n$ with a
\defterm{symplectic form $\gensymplstr$} that satisfies
\begin{equation}\begin{aligned}
  {}&\gensymplstr^\qformT = - \gensymplstr \comma
    \gensymplstr\in\tenssct_2^0\,\phspc \commae\\
  {}&\gensymplstr \qquad \text{is non-degenerate}\commae\\
  {}&\gensymplstr \qquad \text{is closed (i.e. $\grad\gensymplstr = 0$)}\period
\end{aligned}\end{equation}
We can invert it\footnote{
  The dot $\phint$ indicates the contraction in $\tens\,\phspc$.
  }
\begin{equation}
  \gensymplstr^{\dash 1}\phint\gensymplstr
  = - \deltadst_\phspc
\end{equation}
and define a \defterm{canonical vector field} associated with a function $H$ on
$\phspc$
\begin{equation}
  \canvf{H} = (\grad H)\phint
  \gensymplstr^{\dash 1}\period
\end{equation}
This canonical vector field generates a canonical transformation on $\phspc$
that does not change the symplectic structure:
\begin{equation}
  \lieder_{\canvf{H}} \gensymplstr = 0 \period
\end{equation}
We can define the \defterm{Poisson brackets} of functions on $\phspc$ as
\begin{equation}\label{PoisBrDef}
  \{A,B\} = {\canvf{A}}\phint \grad B =
  \canvf{A}\phint\gensymplstr\phint\canvf{B} =
  (\grad A)\phint\gensymplstr^{\dash 1}\phint
  (\grad B) \period
\end{equation}

We have
\begin{gather}
  \frac d{dt} B \defeq \lieder_{\canvf{H}} B = \{H,B\}\commae\\
  [\canvf{A},\canvf{B}] = \canvf{\{A,B\}} \commae
\end{gather}
where $[\:,]$ are the Lie brackets on vector fields. The symplectic structure also
induces a volume element on $\phspc$
\begin{equation}\label{SymMsrDef}
  \symplmsr = (2\pi)^{-n}\,\frac 1{n!}\;
  \underbrace{\abs{\,\gensymplstr\wedge\gensymplstr
  \wedge\dots\wedge\gensymplstr\,}}_{\text{$n$ times}} =
  \Bigl(\Det \frac{\gensymplstr}{2\pi}\Bigr)^{\frac12} \period
\end{equation}

Finally, if we choose coordinates $(x^a, p_a)$ for $a=1,\dots,n$ such that
\begin{equation}
  \gensymplstr = \grad p_a\wedge\grad x^a \commae
\end{equation}
we get
\begin{equation}
  \{ x^a,p_b\} = - \delta_b^a
\end{equation}
and $(x^a, p_a)$ are \defterm{canonical coordinates}.


\section{Tangent and cotangent bundle geometry}
\label{apx:TSpG}

In this appendix we discuss the geometry of tangent bundles of a manifold
$\valspc$. We define \vague{partial derivatives} of observables on these spaces
in a covariant way. We also show that the cotangent bundle has the structure of
a symplectic manifold. We use ${\scriptstyle \aixval{a}}, {\scriptstyle
\aixval{b}},\dots$ as indices for tensors on the manifold $\valspc$ and indices
$\aixbph{A}, \aixbph{B},\dots$ for tensors from tangent spaces of the cotangent
bundle $\phspc = \tens^\dual\,\valspc$.

Functions such as the Lagrangian $L(x,v)$ and the Hamiltonian $H(x,p)$ are
functions on the tangent and cotangent bundle, respectively, of a configuration
space $\valspc$. Because velocities $v$ (or momenta $p$) are vectors
(covectors) from different fibers for different position $x$ ($\tens_x\,\valspc
\neq \tens_y\,\valspc$ for $x\neq y$), we have to be careful to use a partial
derivative with respect of the position $x$. There is no problem with the
definition
\begin{equation}
  \frac{\partial L}{\partial v^{\aixval{a}}}(x,v)\;:\qquad
  \delta v^{\aixval{n}}
  \frac{\partial L}{\partial v^{\aixval{n}}}(x,v)
  = \frac d{d\varepsilon} L(x,v+\varepsilon \delta v) |_{\varepsilon = 0}
\end{equation}
--- a \defterm{partial derivative with $x$ constant} --- but to define the
\defterm{derivative with $v$ constant} we need a connection $\valcnx$ on
$\valspc$
\begin{equation}
  \frac{\valcnx_{\aixval{a}} L}{\partial x}(x,v)\;:\qquad
  \delta x^{\aixval{n}}
  \frac{\valcnx_{\aixval{n}} L}{\partial x}(x,v)
  = \frac d{d\varepsilon} L(x_\varepsilon,v_\varepsilon)
  |_{\varepsilon = 0}\commae
\end{equation}
where $x_\varepsilon$ is a curve starting from $x$ in a direction $\delta x$ and
$v_\varepsilon$ is the parallel transport of $v$ along $x_\varepsilon$ in the
sense of the connection $\valcnx$ (i.e.
$\frac{\valcnx}{d\varepsilon}v_\varepsilon = 0$). Similarly, for a function on
the cotangent bundle,
\begin{equation}\begin{gathered}
  \frac{\partial H}{\partial p_{\aixval{a}}}(x,p)\;:\qquad
  \delta p_{\aixval{n}}
  \frac{\partial H}{\partial p_{\aixval{n}}}(x,p)
  = \frac d{d\varepsilon} H(x,p+\varepsilon \delta p)
  |_{\varepsilon = 0}\commae\\
  \frac{\valcnx_{\aixval{a}} H}{\partial x}(x,p)\;:\qquad
  \delta x^{\aixval{n}}
  \frac{\valcnx_{\aixval{n}} H}{\partial x}(x,p)
  = \frac d{d\varepsilon} H(x_\varepsilon,p_\varepsilon)
  |_{\varepsilon = 0}\commae
\end{gathered}\end{equation}
where again $\frac{\valcnx}{d\varepsilon}p_\varepsilon = 0$.

We want to show that the cotangent bundle $\phspc = \tens^\dual\,\valspc$ has the
structure of a phase space. It will be useful to define a covariant
generalization of \vague{coordinate} vector fields and forms
\begin{equation}\begin{aligned}
  {}&\frac{\partial^{\aixbph{A}}}{\partial p_{\aixval{a}}}&\qquad
  &\text{a vector field on $\phspc$ for which}\quad
  \frac{\partial^{\aixbph{A}}}{\partial p_{\aixval{a}}}
  \grad_{\aixbph{A}}H =
  \frac{\partial H}{\partial p_{\aixval{a}}}\commae\\
  {}&\frac{\valcnx_{\aixval{a}}^{\aixbph{A}}}{\partial x}&\qquad
  &\text{a vector field on $\phspc$ for which}\quad
  \frac{\valcnx_{\aixval{a}}^{\aixbph{A}}}{\partial x}
  \grad_{\aixbph{A}}H =
  \frac{\valcnx_{\aixval{a}} H}{\partial x}\period
\end{aligned}\end{equation}
$\frac{\partial}{\partial p}$ is actually the natural identification of the
vector space $\tens^\dual_x\,\valspc$ with its tangent space
$\tens(\tens^\dual_x\,\valspc)$ and $\frac{\valcnx}{\partial x}$ is the
horizontal shift of the connection $\valcnx$. Form fields dual to these vector fields
\begin{equation}\begin{aligned}
  {}&\Dmap_{\aixbph{A}}^{\aixval{a}}x\comma
  \text{differential of the bundle projection }\;
  x\,:\;\tens\,\valspc\rightarrow\valspc,\;p|_x\rightarrow x\commae\\
  {}&\valcnx_{\aixbph{A}} p_{\aixval{a}} \commae
\end{aligned}\end{equation}
are defined by
\begin{equation}\begin{gathered}\label{ctgdual}
  \frac{\valcnx_{\aixval{a}}^{\aixbph{N}}}{\partial x}\,
  \Dmap_{\aixbph{N}}^{\aixval{b}} x=
  \delta^{\aixval{b}}_{\aixval{a}}\comma
  \frac{\partial^{\aixbph{N}}}{\partial p_{\aixval{b}}}\,
  \valcnx_{\aixbph{N}} p_{\aixval{a}} =
  \delta^{\aixval{b}}_{\aixval{a}}\comma
  \frac{\valcnx_{\aixval{a}}^{\aixbph{N}}}{\partial x}\,
  \valcnx_{\aixbph{N}} p_{\aixval{b}} = 0 \comma
  \frac{\partial^{\aixbph{N}}}{\partial p_{\aixval{a}}}\,
  \Dmap_{\aixbph{N}}^{\aixval{b}} x = 0\commae\\
  \frac{\valcnx_{\aixval{n}}^{\aixbph{A}}}{\partial x}\,
  \Dmap_{\aixbph{B}}^{\aixval{n}} x +
  \frac{\partial^{\aixbph{A}}}{\partial p_{\aixval{n}}}\,
  \valcnx_{\aixbph{B}} p_{\aixval{n}} = \delta^{\aixbph{A}}_{\aixbph{B}}\period
\end{gathered}\end{equation}

Now we can write down the canonical cotangent bundle symplectic form
\begin{gather}
  \gensymplstr_{\aixbph{AB}} =
  \valcnx_{\aixbph{A}} p_{\aixval{a}}\wedge
  \Dmap_{\aixbph{B}}^{\aixval{a}} x =
  \valcnx_{\aixbph{A}} p_{\aixval{a}}\,
  \Dmap_{\aixbph{B}}^{\aixval{a}} x -
  \Dmap_{\aixbph{A}}^{\aixval{a}} x \,
  \valcnx_{\aixbph{B}} p_{\aixval{a}}\commae \label{CanCTBSymplstr}\\
  \gensymplstr^{\dash1\,\aixbph{AB}} =
  \frac{\partial^{\aixbph{A}}}{\partial p_{\aixval{n}}}
  \frac{\valcnx_{\aixval{n}}^{\aixbph{B}}}{\partial x} -
  \frac{\valcnx_{\aixval{n}}^{\aixbph{A}}}{\partial x}
  \frac{\partial^{\aixbph{B}}}{\partial p_{\aixval{n}}}
\end{gather}
and we can even explicitly write the symplectic potential
\begin{equation}
  \gensymplstr_{\aixbph{AB}} = \grad_{\aixbph{A}}\theta_{\aixbph{B}} \comma
  \theta_{\aixbph{A}} = p_{\aixval{n}} \Dmap_{\aixbph{A}}^{\aixval{n}} x \period
\end{equation}
The canonical vector fields and Poisson brackets are
\begin{gather}
  \canvf[\aixbph{A}]{F} =
  \frac{\partial F}{\partial p_{\aixval{n}}}
  \frac{\valcnx_{\aixval{n}}^{\aixbph{A}}}{\partial x} -
  \frac{\valcnx_{\aixval{n}} F}{\partial x}
  \frac{\partial^{\aixbph{A}}}{\partial p_{\aixval{n}}}\commae\\
  \{A,B\} =
  \frac{\partial A}{\partial p_{\aixval{n}}}
  \frac{\valcnx_{\aixval{n}} B}{\partial x} -
  \frac{\valcnx_{\aixval{n}} A}{\partial x}
  \frac{\partial B}{\partial p_{\aixval{n}}}\period
\end{gather}

If we change the connection to another one,
\begin{equation}\begin{gathered}
  \tilde\valcnx = \valcnx \oplus \Gamma\commae\\
  \tilde\valcnx_{\aixval{a}} a^{\aixval{b}} =
  \valcnx_{\aixval{a}} a^{\aixval{b}} +
  \Gamma_{\aixval{an}}^{\aixval{b}} a^{\aixval{n}} \comma
  \tilde\valcnx_{\aixval{a}} p_{\aixval{b}} =
  \valcnx_{\aixval{a}} p_{\aixval{b}} -
  \Gamma_{\aixval{ab}}^{\aixval{n}} p_{\aixval{n}} \comma
\end{gathered}\end{equation}
we get
\begin{equation}\begin{gathered}
  \frac{\tilde\valcnx_{\aixval{a}}^{\aixbph{A}}}{\partial x}(x,p) =
  \frac{\valcnx_{\aixval{a}}^{\aixbph{A}}}{\partial x}(x,p) +
  p_{\aixval{r}} \Gamma_{\aixval{as}}^{\aixval{r}}(x)
  \frac{\partial^{\aixbph{A}}}{\partial p_{\aixval{s}}}(x,p)\commae\\
  \tilde\valcnx_{\aixbph{A}} p_{\aixval{a}}(x,p) =
  \valcnx_{\aixbph{A}} p_{\aixval{a}}(x,p) -
  p_{\aixval{r}} \Gamma_{\aixval{as}}^{\aixval{r}}(x)
  \Dmap_{\aixbph{A}}^{\aixval{s}}x(x,p)\period
\end{gathered}\end{equation}
By straightforward calculations, we can check that the quantities
$\gensymplstr$, $\theta$, $\canvf{F}$, and $\{\:,\}$ do not depend on the
choice of the connection.

Finally, coordinates $x^{\ixval{a}}$ on $\valspc$ generate coordinates
$(x^{\ixval{a}}, p_{\ixval{a}})$ on $\phspc$ by
\begin{equation}
  p_{\ixval{a}} = p_{\aixval{a}}
  \frac{\partial^{\aixval{a}}}{\partial x^{\ixval{a}}}
\end{equation}
and they define the coordinate connection ${\partial}$ on $\valspc$ for which
\begin{equation}
  {\partial} \grad x^{\ixval{a}} = 0 \comma
  {\partial} \frac{\partial}{\partial x^{\ixval{a}}} = 0
  \qquad\text{for ${\ixval{a}} = 1,2,\dots,n$}\period
\end{equation}
Using this connection and expressing everything in coordinates, we get the
standard relations \cite{Arnold:book,Frankel:book}
\begin{gather}
  \gensymplstr_{\aixbph{AB}} =
  \grad_{\aixbph{A}}p_{\ixval{n}}\wedge
  \grad_{\aixbph{B}} x^{\ixval{n}}\comma
  \theta_{\aixbph{A}} = p_{\ixval{n}}
  \grad_{\aixbph{A}} x^{\ixval{n}}\commae\\
  \canvf[\aixbph{A}]{F} =
  \frac{\partial F}{\partial p_{\ixval{n}}}
  \frac{\partial^{\aixbph{A}}}{\partial x^{\ixval{n}}} -
  \frac{\partial F}{\partial x^{\ixval{n}}}
  \frac{\partial^{\aixbph{A}}}{\partial p_{\ixval{n}}}\commae\\
  \{A,B\} =
  \frac{\partial A}{\partial p_{\ixval{n}}}
  \frac{\partial B}{\partial x^{\ixval{n}}} -
  \frac{\partial A}{\partial x^{\ixval{n}}}
  \frac{\partial B}{\partial p_{\ixval{n}}}\commae\\
  \{x^{\ixval{a}},p_{\ixval{b}}\} =
  - \delta^{\ixval{a}}_{\ixval{b}}\period
\end{gather}


\section{Densities on a Manifold}
\label{apx:Dens}

In this appendix we shortly review the definition of densities on a manifold
and some operations with them.

On any manifold $M$ we can define a vector bundle of tangent densities of a
weight $\alpha$ that we denote $\denst{\realn}^\alpha M$ and
$\denst{\complexn}^\alpha M$ if the densities are real and complex,
respectively. The space of sections we denote $\dnstsct^\alpha M$. The standard
fiber of these bundles is the vector space of real or complex numbers. As for
any tangent bundle, the density bundle can be defined by a \defterm{coordinate
map} from the space of frames of the tangent vector space $\tens\,M$ to the
standard fiber $\realn$ (or $\complexn$). The coordinate map has to be a
representation of the linear group acting on the space of frames. The
coordinate map $e_a\rightarrow \mu[e_a]$ tells us by what factor is a density
$\mu$ different from the coordinate density $\epsilon$ given by the base $e_a$
\begin{equation}
  \mu[e_a] = \mu\, \epsilon^{\dash 1} \comma \epsilon[e_a] = 1 \period
\end{equation}
For densities of weight $\alpha$, the coordinate map is a representation of the
linear group of the following type:
\begin{equation}\label{DensTrProp}
  \mu[A_a^b e_b] = \abs{\det A}^\alpha\,\mu[e_a] \period
\end{equation}
Clearly, we can define complex densities even for a complex weight.

Besides the linear operation, we can also define the multiplication and
constant powers of densities. These operations map densities of some weight to
densities of a different weight. Let us note that a complex conjugation maps
the densities of weight $\alpha$ to densities of weight $\alpha^\cxc$ and
therefore \vague{the absolute value} $\abs{\mu}=(\mu \mu^\cxc)^{\frac12}$
belongs to densities of weight $\re\alpha$.

Densities of weight $1$ are called \defterm{volume elements} because they can
be integrated. Let $\mu$ be a volume element, then we define locally
\begin{equation}
  \int_\Omega \mu =
  \int_{x^a(\Omega)} \mu\Bigl[\frac{\partial}{\partial x^a}\Bigr]\; \dvol[n]{x}\commae
\end{equation}
where $x^a$ are arbitrary coordinates. On the right hand side the usual
coordinate integration is understood. Extension to domains not covered by one
coordinate system is done by standard methods \cite{HawkingEllis:book}. The
consistency of this definition (independence of a choice of coordinates)
follows from the transformation properties \eqref{DensTrProp} with $\alpha=1$.

A metric $g$ or a symplectic form $\gensymplstr$ (see Appendix~\ref{apx:SySp})
define canonical volume elements:
\begin{equation}
  \vol{g} = \abs{\Det g}^{\frac12}\comma
  \symplmsr = \abs{\Det \frac{\symplstr}{2\pi}}^{\frac12}\commae
\end{equation}
where the operation $\Det : \tens^0_2\,M\rightarrow \denst{\realn}^1 M$ is
defined by
\begin{equation}
  (\Det g)[e_a] = \det g_{ab}\period
\end{equation}

We can also define a map from the space of totally antisymmetric forms to the
space of volume elements $\sigma\rightarrow\abs{\sigma}$, such that
\begin{equation}
  \abs{\sigma}[e_a] = \abs{\sigma_{1\dots n}}\commae
\end{equation}
$n$ beeing the dimension of the manifold $M$. It is easy to check that
\eqref{DensTrProp} is  satisfied.

Finally, let us note that the delta distribution $\delta(x|y)$ can be
understood as a bi-distribution with a density character in both arguments and
of weights $\alpha$, $\beta$ such that $\alpha+\beta=1$, i.e., it is a
functional on test densities $\mu$, $\nu$ of weights $1-\alpha$ and $1-\beta$:
\begin{equation}
  \int \mu(x) \nu(y) \delta(x|y) =
  \int \mu\nu\period
\end{equation}
If we want to define the \vague{ordinary} delta function that is not a density
in any of its arguments, we have to normalize it to some volume
element\footnote{
  For distributions on $\realn^n$, one usually chooses the canonical volume
  element $\dvol[n]{x}$.
  }
$\mu$. We call such a distribution $(\mu^{\dash 1}\delta)$. It acts on test
densities $\varphi$, $\psi$ of weight $1$:
\begin{equation}
  \int \varphi(x) \psi(y)\,(\mu^{\dash 1}\delta)(x|y)
  = \int \varphi \psi \mu^{\dash 1}\period
\end{equation}
Similarly, for any smooth density $f$ we can define a bi-distribution
$(f\delta)$.

\end{document}